\documentclass[aps,nofootinbib,superscriptaddress,10pt,floatfix]{revtex4}
\usepackage{mathrsfs}
\usepackage{graphicx}
\usepackage{amsmath,amssymb}
\usepackage{hyperref}
\usepackage{braket}
\usepackage{subfigure}
\usepackage{subfloat}
\usepackage{float}
\usepackage[usenames]{color}
\usepackage[normalem]{ulem}

\hypersetup{
    colorlinks=true,
    linkcolor=red,
    citecolor=blue,
}

\allowdisplaybreaks

\def\be{\begin{equation}}
\def\ee{\end{equation}}
\def\ba{\begin{eqnarray}}
\def\ea{\end{eqnarray}}

\def\mn{\mu\nu}

\begin{document}
	
\title{On validity of the quasi-static approximation in scalar-tensor theories}
	
\author{Seyed Hamidreza Mirpoorian}
\email[]{smirpoor@sfu.ca}

\author{Zhuangfei Wang}
\email[]{zhuangfei_wang@sfu.ca}

\author{Levon Pogosian}
\email[]{levon@sfu.ca}
\affiliation{Department of Physics, Simon Fraser University, Burnaby, BC, V5A 1S6, Canada}

\begin{abstract}
The discovery of cosmic acceleration motivated extensive studies of dynamical dark energy and modified gravity models. Of particular interest are the scalar-tensor theories, with a scalar field dark energy non-minimally coupled to matter. Cosmological constraints on these models often employ the quasi-static approximation (QSA), in which the dynamics of the scalar field perturbations is proportional to the perturbation in the matter density. Using the QSA simplifies the physical interpretation of the phenomenology of scalar-tensor theories, and results in substantial savings of computing time when deriving parameter constraints. Focusing on the symmetron model, which is a well-motivated scalar-tensor theory with a screening mechanism, we compare the exact solution of the linearly perturbed field equations to those obtained under the QSA and identify the range of the model parameters for which the QSA is valid. We find that the evolution of background scalar field is most important, namely, whether it is dominated by the Hubble friction or the scalar field potential. This helps us derive a criterion for the symmetron model, but same argument can be applied to other scalar-tensor theories of generalized Brans-Dicke type. We consider two scenarios, one where the scalar field is only coupled to dark matter and where it couples to all of the matter.
\end{abstract}
	
\maketitle
	
\section{Introduction}

The $\Lambda$ Cold Dark Matter ($\Lambda$CDM) model provides an acceptable fit to the data, yet many other dark energy (DE) and modified gravity (MG) models have been proposed and studied in the literature~\cite{Copeland:2006wr,Silvestri:2009hh,Clifton:2011jh, Joyce:2014kja} with the aim to explain the observed cosmic acceleration~\cite{Perlmutter:1998np,Riess:1998cb}. The primary motivation for these studies comes from the extreme fine-tuning required to reconcile the small observed value of $\Lambda$ with the large vacuum energy density predicted by particle physics~\cite{Weinberg:1988cp,Burgess:2013ara}. The necessity to postulate yet another dark component, CDM, the evidence for which is purely gravitational, further adds to the interest in possible extensions of General Relativity (GR).

Modifications of GR typically involve new degrees of freedom~\cite{Clifton:2011jh}, such as a scalar field which can, along with being a source of dynamical DE, mediate a force between matter particles often referred to as the "fifth force". There are very tight constraints on the presence of such a fifth force acting on matter that is part of the standard model of particle physics~\cite{Will:2014kxa}. Hence, any modified gravity model with a scalar field non-minimally coupled to all matter that is designed to have cosmological implications must also include a screening mechanism that would suppress the fifth force inside the Solar System~\cite{Vainshtein:1972sx,Damour:1994zq,Khoury:2003aq,Hinterbichler:2010es,Joyce:2014kja}. On the other hand, if the scalar field coupled only to DM, then the Solar System tests of gravity would not apply and the screening would not be necessary. While it is unnatural for a light scalar field playing the role of DE not to couple to all matter~\cite{Carroll:1998zi}, once one allows for the existence of a dark sector, one should allow for the possibility that DM and DE have their own special interaction -- a possibility extensively studied in the literature~\cite{Amendola:1999er,PhysRevLett.64.123,delaMacorra:2002tk,Chimento:2003iea}. Ultimately, we would like to use observational data to differentiate between the dark-matter-only and all-matter coupled DE, although this can be challenging in practice~\cite{Bonvin:2022tii}.

Deriving cosmological constraints on scalar-tensor theories often rely on the quasi-static approximation (QSA), where the scalar field is assumed to be tracking the evolution of the matter fluid to which it is coupled. Using the QSA not only helps to simplify the numerical implementation, potentially saving substantial computing time required to resolve rapid scalar field oscillations, but also makes it possible to relate parameters of the scalar-tensor theory to the phenomenological parameters, such as the effective Newton's constant and the gravitational slip, used to constrain modifications of GR on cosmological scales with the help of publicly available codes such as MGCAMB and MGCLASS~\cite{Zhao:2008bn,Hojjati:2011ix,Zucca:2019xhg,Sakr:2021ylx}. Understanding the range of applicability of the QSA is, therefore, important for the practical implementation and meaningful interpretation of model-agnostic constraints on scalar-tensor theories.

In this paper, we consider the symmetron model~\cite{Hinterbichler:2010es,Hinterbichler:2011ca}, which is a well-motivated scalar-tensor theory with a screening mechanism and a renormalizable potential. The potential of the symmetron field acts as the effective cosmological constant which causes the accelerated expansion of the universe at late times. The matter particles feel the presence of a fifth force at late time, due to the non-minimal coupling with the scalar field. When matter density is higher than a certain critical value, the coupling vanishes and one recovers GR. This ensures that GR is recovered at early times and in dense regions such as our galaxy.

In what follows, we provide a comprehensive analysis of the evolution of the background cosmology and linear perturbations in the symmetron model, with and without using the QSA. We also separately consider the cases of the symmetron coupled only to DM and to all mater. We find that in both cases the QSA provides an accurate solution when the combination of the Compton wavelength and the redshift of the symmetry breaking transition is relatively small, but breaks down when this combination is large. Generally, the QSA holds when the background solution for the scalar field evolves sufficiently slowly over a period of oscillation of the scalar field perturbation, set by the inverse of the Compton wavelength. This paves the way to using MGCAMB and MGCLASS for deriving observational constraints on the parameters of the symmetron model and other scalar-tensor theories.
	
\section{The quasi-static approximation for a scalar field coupled to dark matter and all matter} 
\label{s:dm and de}
	
In this Section, we present the equations of motion of a cosmological theory that includes a scalar field conformally coupled to matter, starting with the case where the scalar field is coupled only to DM. Throughout this paper, $g_{\mn}$ denotes the \emph{baryon frame} metric whose geodesics are followed by the Standard Model (SM) particles, or ``baryons'', since the observable quantities are measured in this frame. We use the metric signature $(-,+,+,+)$.

Evolving the scalar field equations can be time consuming due to rapid oscillations, making it impractical for applications requiring many repeated runs, such as MCMC parameter constraints. Instead, in some circumstances, one can use the QSA in which the scalar field perturbation is algebraically related to the matter density contrast. The QSA involves two assumptions: (1) that the Fourier modes of interest are sub-horizon, $k/\mathcal{H} \gg 1$, and (2) that the time derivatives of the metric and the scalar field are much smaller than their spatial derivatives. In $\Lambda$CDM, the sub-horizon condition automatically implies the smallness of the time derivatives of the linear metric perturbations, as their growth rate becomes comparable to spatial derivatives only for perturbations on near-Hubble scales. In scalar-tensor theories, however, the scalar field has its own dynamics, and the first assumption no longer implies the second. Moreover, the time-derivatives of the scalar field and the metric are not necessarily small. However, the QSA can still be a good approximation if neglecting the time-derivatives has no impact on observables. For instance, rapid time-oscillations of the scalar field around a central position may simply average out in their contribution to the observed matter power spectrum~\cite{Hojjati:2012rf}.

In what follows, we present the exact and the QSA form of the relevant perturbation equations with more details provided in the Appendix.

\subsection{Scalar field coupled to dark matter} \label{ss:cdmonly}
	
Let us consider the following action:
	\begin{equation}
		\label{eq:action}
		S = \int \mathrm{d}^4 x \left\{ \sqrt{-g} \left[ \frac{\mathcal{R}}{16\pi G} - \frac{1}{2} \partial_{\mu} \phi \, \partial^{\mu} \phi - V(\phi) \right] + \mathcal{L}_{\mathrm{dm}} (\psi_\mathrm{dm}, \tilde{g}_{\mu \nu}) + \mathcal{L}_{\mathrm{SM}} (\psi_\mathrm{m}, g_{\mu \nu}) \right\},
	\end{equation}
with the scalar field $\phi$ and the potential $V$ measured in $M_\mathrm{Pl}$ and $M_{\mathrm{Pl}}^2 \mathrm{Mpc}^{-2}$ units, respectively, where $M_\mathrm{Pl} \equiv (8\pi G)^{-1/2}$ is the reduced Plank mass ($\mathrm{Mpc}$ denotes megaparsecs). In the above, $\mathcal{L}_{\mathrm{SM}}$ is the Lagrangian density of the standard model of particle physics, describing the known baryons, leptons and gauge bosons, and $\mathcal{L}_{\mathrm{dm}}$ is the Lagrangian of DM particles. The metric $g_{\mu\nu}$ determines the curvature scalar $\mathcal{R}$ and defines the \emph{baryon frame} in which the SM particles follow the geodesics of $g_{\mu\nu}$. The DM particles follow the geodesics of $\tilde{g}_{\mu \nu}$, related to $g_{\mu\nu}$ through a conformal transformation
	\begin{equation}
		\label{eq:conformal trans}
		\tilde{g}_{\mu \nu} = A^2 (\phi) g_{\mu\nu} \, ,
	\end{equation}
where $A(\phi)$ is a coupling function. When the coupling function is non-constant, $A,_{\phi} \ne 0$, the scalar field mediates a fifth force between DM particles. In this DM-only coupled case, the form of the Einstein-Hilbert action is unmodified in the baryon frame. 

Next, we consider the perturbed flat Friedmann-Lemaitre-Robertson-Walker (FLRW) metric in the \emph{Conformal Newtonian gauge},
\begin{equation}
	\label{eq:metric per}
	ds^2 = a^2(\tau) \left[ -(1 + 2\Psi) d\tau^2 + (1-2\Phi) d \vec{x}^2 \right],
\end{equation}
where $a$ is the scale factor and $\tau$ is the conformal time. The two scalar potentials, $\Psi$ and $\Phi$, characterize the Newtonian gravitational potential and the curvature perturbations, respectively. We write the scalar field as a sum of the homogeneous (background) part and a perturbation which depends on space and time,
\begin{equation}
	\label{eq:de parts}
	\phi(\vec{x}, \tau) = \phi^{(0)} (\tau) + \delta \phi (\vec{x}, \tau),
\end{equation}
and, similarly, split the stress-energy components into their background and perturbation parts.

The complete set of equations for the background quantities and linear metric and matter perturbations, before and after applying the QSA, is given by

\hspace{-0.4cm}

\begin{tabular}{c c}
	\bf{Exact} & \bf{QSA} \\
	\parbox[t]{0.52\linewidth}{
		\begin{subequations} \label{eq:exact_cdmonly}
			\renewcommand{\theequation}{\theparentequation.\arabic{equation}}
			\begin{align}
				&\mathcal{H}^2 = \frac{8\pi G a^2}{3} 
				\left(
				\frac{\dot{\phi}^2}{2 a^2} + V(\phi) + \rho_{\mathrm{dm}} + \rho_{b} + \rho_{\gamma} + \rho_{\nu} 
				\right) \label{eq:friedmann_cdm} \\
				&\ddot{\phi} + 2 \mathcal{H} \dot{\phi} = -a^2 V'_{\rm eff} \label{eq:deeom_cdm} \\
				&\dot{\rho}_{\mathrm{dm}} + 3 \mathcal{H} \rho_{\mathrm{dm}} = \beta \dot{\phi} \rho_{\mathrm{dm}} \label{eq:dmeom_cdm} \\
				\begin{split}
					&k^2 \Phi + 3\mathcal{H} (\dot{\Phi} + \mathcal{H} \Psi) = -4 \pi Ga^2 
					\Biggl( 
					{\sum\limits_{i} } \rho_{i} \delta_{i} + V,_{\phi} \delta{\phi} \\[-3ex]
					&\quad+ \frac{\dot{\phi} \delta{\dot{\phi}}}{a^2} - \Psi \frac{\dot{\phi}^2}{a^2} 
					\Biggr) \label{eq:cphi_cdm}
				\end{split} \\
				&k^2 (\Phi - \Psi) = 16 \pi G a^2 ( \rho_{\gamma} \sigma_{\gamma} + \rho_{\nu} \sigma_{\nu}) \label{eq:shear_cdm} \\
				&\dot{\delta}_b + \theta_b = 3 \dot{\Phi} \label{eq:deltadotb_cdm} \\
				&\dot{\theta}_b + \mathcal{H} \theta_b = k^2 \Psi
				+ c_s^2 k^2 \delta_b + \frac{\dot{\tau}_c}{R} 
				\left( \theta_{\gamma} - \theta_b \right) \label{thetadotb_cdm} \\
				&\dot{\delta}_{\mathrm{dm}} =
				- \theta_{\mathrm{dm}} + 3\dot{\Phi} + \beta \delta \dot{\phi} 
				+ \dot{\beta} \delta \phi \label{eq:deltadotdm_cdm} \\
				&\dot{\theta}_{\mathrm{dm}} = - ( \mathcal{H}
				+ \beta \dot{\phi}) \theta_{\mathrm{dm}}  + k^2 \left( \Psi + \beta \delta \phi \right) \label{eq:thetadotdm_cdm} \\
				\begin{split}
					&\delta \ddot{\phi} + 2 \mathcal{H} \delta \dot{\phi} + k^2 \delta \phi = 
					-a^2 (
					2 \Psi V'_{\rm eff} +
					V''_{\rm eff} \delta \phi \\
					&+ \beta \rho_{\mathrm{dm}} \delta_{\mathrm{dm}}
					)
					+ \dot{\phi} ( \dot{\Psi} + 3\dot{\Phi}) \label{eq:ddeeom_cdm}
				\end{split}
			\end{align}
		\end{subequations}
	} &
	\parbox[t]{0.453\linewidth}{
		\begin{subequations} \label{eq:qsa_cdmonly}
			\renewcommand{\theequation}{\theparentequation.\arabic{equation}}
			\begin{align}
				&\mathcal{H}^2 = \frac{8\pi G a^2}{3} 
				\bigg(
				\rho_{\mathrm{dm}} + \rho_{b} + \rho_{\gamma} + \rho_{\nu} + \rho_{\Lambda}
				\bigg) \label{eq:fiedmann_lcdm} \\
				&\phi = \phi_{\rm min} \\
				&\rho_{\mathrm{dm}} = \frac{A}{A_0} \frac{\rho_{\mathrm{dm}}^{(0)}}{a^3} \label{eq:rhodm_qsacdm} \\[2ex]
				&k^2 \Phi = -4 \pi Ga^2 {\sum\limits_{i} } \rho_{i} \delta_{i} \label{eq:cphi_qsacdm} \\[2.5ex]
				&k^2 (\Phi - \Psi) = 16 \pi G a^2 ( \rho_{\gamma} \sigma_{\gamma} + \rho_{\nu} \sigma_{\nu}) \label{eq:shear_qsacdm} \\
				&\dot{\delta}_b + \theta_b = 0 \label{eq:deltadotb_qsacdm} \\[1ex]
				&\dot{\theta}_b + \mathcal{H} \theta_b = k^2 \Psi \label{eq:thetadotb_qsacdm} \\[1ex]
				&\dot{\delta}_{\mathrm{dm}} + \theta_{\mathrm{dm}} = 0 \label{eq:deltadotdm_qsacdm} \\
				&\dot{\theta}_{\mathrm{dm}} +  \mathcal{H} \theta_{\mathrm{dm}}  
				= k^2 \left( \Psi + \beta \delta \phi \right) \label{eq:thetadotdm_cdmqsa} \\[1ex]
				&\delta \phi = - \frac{\beta \rho_{\mathrm{dm}} \delta_{\mathrm{dm}}}{m^2 + k^2/a^2} \label{eq:deltaphi_qsacdm} 
			\end{align}
		\end{subequations}
	}\\
\end{tabular}
where $\mathcal{H} = \dot{a}/a$, with the dot denoting the derivative with respect to $\tau$, $\rho_{\mathrm{dm}}$, $\rho_b$, $\rho_{\gamma}$, and $\rho_{\nu}$ are the DM, baryon, photon, and neutrino energy densities, respectively, while $\delta_i$, $\theta_i$, and $\sigma_i$ are the energy density fluctuations, divergence of fluid velocity, and anisotropic stress, respectively, following the notation of \cite{Ma:1995ey}. The coupling function $\beta = A,_{\phi} /A$ determines the strength of the DM-scalar interaction, $c_s$ is the baryon sound speed, and $\dot{\tau}_c$ is the differential optical depth which is important before and around the epoch of recombination, where the photons and baryons are tightly coupled. The effective potential $V_{\rm eff}$ is defined via its derivative appearing on the right hand side of Eq.~(\ref{eq:deeom_cdm}),
\begin{align}
	V'_{\mathrm{eff}} &= V,_{\phi} + \beta \rho_{\mathrm{dm}}, \label{eq: vprimeeff}
\end{align}
and the effective scalar field mass appearing in Eq.~\eqref{eq:deltaphi_qsacdm} is $m^2 = V''_{\mathrm{eff}}(\phi_{\rm min})$, where $\phi_{\rm min}$ is the scalar field at the minimum of $V_{\rm eff}$. The continuity equation for DM, Eq.~\eqref{eq:dmeom_cdm}, can be analytically integrated, giving the solution provided in Eq.~\eqref{eq:rhodm_qsacdm}, where $A_0$ and $\rho_{\mathrm{dm}}^{(0)}$ are the corresponding values of the coupling function and DM energy density at present time.

Under the QSA, where the Fourier modes of interest are sub-horizon, the time derivatives of both the metric and the scalar field perturbations are assumed to be negligible compared to their spatial derivatives and removed from Eqs.~\eqref{eq:cphi_cdm}, \eqref{eq:deltadotb_cdm}, \eqref{eq:deltadotdm_cdm}, and \eqref{eq:ddeeom_cdm}.
At the background level, the QSA implies that the scalar field evolves slowly, remaining at the minimum of the slowly changing effective potential with no oscillations around the minimum. In this approach, one eliminates the derivative of the background scalar field in equations where it appears, and obtains an algebraic relation between the background scalar field and the matter density. This is justified as long as $\dot{\phi} \ll \mathcal{H}$. One should, however, keep in mind that this is not a good approximation in models in which $\beta \dot{\phi} \sim {\cal H}$~\cite{Baldi:2010pq}. If the background field is at the minimum of the effective potential, we have $V'_{\mathrm{eff}} = 0$,  and the DM density perturbation is then also algebraically related to the scalar field perturbation $\delta\phi$ via Eq.~\eqref{eq:deltaphi_qsacdm}. Also, the contribution of the scalar field density perturbations, $V,_{\phi} \delta{\phi} $, to the Poisson equation is negligible compared to the matter density, hence we have omitted it in Eq.~(\ref{eq:cphi_qsacdm}). The complete set of equations that we numerically evolve to compare the exact solution to those in the QSA are given in Appendix \ref{ss:cdm_only}, along with additional details on how they are derived.
	
\subsection{Scalar field coupled to all matter} 
\label{ss:all matter}

The action in the case of a scalar field conformally coupled to all matter, when written in the \emph{Einstein frame} (in which the gravitational part of action has its usual form) is similar to the one in Eq.~\eqref{eq:action}, except now both normal matter and DM follow the geodesics of $\tilde{g}_{\mu \nu}$. In the Einstein frame, the baryon continuity and Euler equations acquire the explicit $\phi$-dependent terms previously present only in the DM equations. Hence, the baryon and DM perturbations have similar equations of motion, with the additional coupling to photons for the baryons. In our numerical implementation, we evolve the equations in the Einstein frame and use the conformal transformation to convert to the baryon frame $g_{\mu \nu}$ when interpreting the results, since it is the frame in which the observational measurements are made.
The action in the baryon frame takes the form
\begin{equation}
	\label{eq:action_total}
	{S} = \int \mathrm{d}^4 x \left\{ \sqrt{-{g}} \left[ A^{-2}(\phi) \frac{\mathcal{R}}{16\pi G} - \frac{1}{2} \partial_{\mu} {\phi} \, \partial^{\mu} {\phi} - {V}({\phi}) \right] + \mathcal{L}_{\mathrm{SM+dm}} (\psi_\mathrm{SM+dm}, g_{\mu \nu}) \right\},
\end{equation}
and the baryon frame stress-energy is
	\begin{equation}
		\label{eq:baryon energy tensor total}
		T_{\mu\nu} = \frac{-2}{\sqrt{-g}} \frac{\delta \mathcal{L}}{\delta g^{\mu\nu}} = A^{-2} (\phi) \tilde{T}_{\mu\nu}.
	\end{equation}
The baryon frame QSA equations that are modified compared to their counterparts in the DM-only coupled case of Eqs.~\eqref{eq:qsa_cdmonly} are 
\begin{subequations} \label{eq:qsa_all}
	\renewcommand{\theequation}{\theparentequation.\arabic{equation}}
	\begin{align}
	&{\rho}_{\mathrm{dm}} = \frac{{\rho}_{\mathrm{dm}}^{(0)}}{{a}^3}, \label{eq:rhodm_qsaall} \\
	&k^2 {\Phi} = -4 \pi G{a}^2 A^2 {\sum\limits_{i} } \rho_{i} \delta_{i} - \beta k^2 \delta \phi, \label{eq:cphi_qsaall} \\
	&k^2 ({\Phi} - {\Psi}) = -2 \beta k^2 \delta \phi, \label{eq:shear_qsaall} \\
	&\dot{{\theta}}_{\mathrm{dm}} +  \mathcal{{H}} {\theta}_{\mathrm{dm}}  
	= k^2  {\Psi}, \label{eq:thetadotdm_qsaall} \\
	&\delta {\phi} = - \frac{{\beta} \left( {\rho}_{\mathrm{dm}} {\delta}_{\mathrm{dm}} + {\rho}_{b} {\delta}_{b} \right)}{{m}^2 + k^2/{a}^2}, \label{eq:deltaphi_qsaall}
	\end{align}
	\end{subequations}
where $m^2 = V''_{\mathrm{eff}}(\phi_{\rm min})$ with $V'_{\mathrm{eff}} = V,_{\phi} + \beta (\rho_{\mathrm{dm}} + \rho_{b})$.
The full set of equations and the conversion between baryon and Einstein frames for all variables are provided in Appendix \ref{app:trans_frames}.

The last term on the right-hand side of \eqref{eq:cphi_qsaall} arises from the conformal transformation of the metric from the Einstein to Jordan frame and describes the effect of the fifth force mediated by the scalar field. In the Einstein frame, the equivalent contribution of the fifth force manifests as an additional force term on the right-hand side of the Euler equation ~\eqref{eq:thetadotdm_cdmqsa}. It is important to note that both \eqref{eq:cphi_qsacdm} and \eqref{eq:cphi_qsaall} are intended to encompass the scalar field density perturbation on the right-hand side in the sum over all species. However, their effect is negligible on sub-horizon scales due to the fact that the sound speed of the propagating degree of freedom is equal to the speed of light $(c_s = c)$, hence the scalar field density contribution is dropped under the QSA.

There are two notable differences between Eqs.~\eqref{eq:qsa_cdmonly} and \eqref{eq:qsa_all}. One is that $\delta \phi$, which is the fifth force potential, is now sourced by both DM and baryon inhomogeneities, implying a larger enhancement of the growth rate. The other important distinction is that the two gravitational potentials, $\Phi$ and $\Psi$, are different, developing the so-called ``gravitational slip''~\cite{Daniel:2008et}. As a result, the effective gravitational couplings affecting relativistic particles, which follow the geodesics of $\Phi+\Psi$, and non-relativistic matter, which follows $\Psi$, become different, making it possible, in principle, to distinguish between the DM-only and all-matter coupling. However, confirming this in practice is challenging, as it is difficult to find observables that measure the true $\Psi$~\cite{Bonvin:2022tii}.

A commonly used practical way to model cosmological perturbations in modified gravity is to introduce the phenomenological functions $\mu, \gamma$, and $\Sigma$, which parameterize the deviations of Eqs.~\eqref{eq:cphi_qsaall} and \eqref{eq:shear_qsaall} from their $\mathrm{\Lambda CDM}$ forms.
These functions are identified by
\begin{subequations}
	\label{eq:phenom}
	\begin{align}
		k^2 \Psi &= -4\pi G a^2 \mu ( a, k ) \rho \Delta, \label{eq:phenom_mu}
		\\
		\Phi &= \gamma(a, k) \Psi, \label{eq:phenom_gamma}
		\\
		k^2 \left( \Phi + \Psi \right) &= -8\pi G a^2  \Sigma(a, k)  \rho \Delta, \label{eq:phenom_sigma}
	\end{align}
\end{subequations}
where $\rho \Delta \equiv \sum \rho_{i} \Delta_{i}$, with $\Delta_{i} = \delta_{i} + 3 \mathcal{{H}} v_i/k$ being the comoving density contrast, and $v_i$ is the irrotational component of peculiar velocity. Under the QSA, $\Delta_{i} = \delta_{i}$. These phenomenological functions are also known as $G^{matter}=\mu(a,k)G$ and $G^{light}= \Sigma(a,k)G$. Note that the three functions are related via $\Sigma=\mu(\gamma+1)/2$. 
Combining Eqs.~\eqref{eq:cphi_qsaall}, \eqref{eq:shear_qsaall} and \eqref{eq:deltaphi_qsaall}, and considering the fact that ${V}'_{\rm eff} = 0$ in QSA implies ${V},_{{\phi}} = -{\beta} ({\rho}_{b} + {\rho}_{\mathrm{dm}})$, one finds the expressions for the phenomenological functions to be (see also \cite{Brax:2011aw,Brax:2012gr})
\begin{subequations}
	\label{eq:phenom_jordan}
	\begin{align}
		\mu(a, k) &= A^2 \left( 1 + \frac{2k^2 \beta^2}{k^2 + m^2 a^2} \right), \label{eq:phenom_mu_j}
		\\
		\gamma(a, k) &= 1 - \frac{2}{1 + \left( k^2 + m^2 a^2 \right) / 2k^2 \beta^2}, \label{eq:phenom_gamma_j}
		\\
		\Sigma(a, k) &= A^2. \label{eq:phenom_sigma_j}
	\end{align}
\end{subequations}

\section{The Symmetron example} \label{s:symmetron_example}

In what follows, we introduce the symmetron model in the CDM-coupled case, with all main features being the same for the all-matter coupled scalar field. In this model, the potential and the coupling function are given by~\cite{Hinterbichler:2010es}
		\begin{align}
		V(\phi) &= V_0 - \frac{1}{2} \mu^2 \phi^2 + \frac{1}{4} \lambda \phi^{4}, \label{eq: scalarfield potential}
		\\
		A(\phi) &= 1 + \frac{1}{2} \left(\frac{\phi}{M}\right)^2, \label{eq:coupling}
		\end{align}
where $V_0$ is set by the present value of the DE density, and $(\mu , M, \lambda)$ are three parameters whose values are discussed later in this section.
Using the DM energy density given by Eq.~\eqref{eq:rhodm_qsacdm} in Eq.~\eqref{eq: vprimeeff} and integrating allows us to write the effective potential as
	\begin{equation}
		V_{\mathrm{eff}} = V(\phi) + \frac{A}{A_0} \frac{\rho_{\mathrm{dm}}^{(0)}}{a^3},
		\label{eq: veff}
	\end{equation}
where $A_0 = 1 + \phi_0^2/2M^2$ is the present value of the coupling function. Putting together Eqs.~\eqref{eq: scalarfield potential}, \eqref{eq:coupling} and \eqref{eq: veff} allows us to write the effective potential in the symmetron model as
	\begin{equation}
		\begin{split}
			V_{\mathrm{eff}} &= 
			V_0 + \frac{\rho_{\mathrm{dm}}^{(0)}}{A_0 a^3}
			+ \frac{\phi^2}{2M^2} \left[ - \mu^2 M^2 + \frac{\rho_{\mathrm{dm}}^{(0)}}{A_0 a^3} \right] + \frac{1}{4} \lambda \phi^4.
			\label{eq:veff symmetron}
		\end{split}
	\end{equation}
The effective potential changes its shape depending on the DM background density. At low matter densities, the effective potential has two minima with $\phi \ne 0$, hence $A \ne 1$ and DE and DM are non-minimally coupled to each other. However, at high densities corresponding to the early universe or highly clustered regions such as the Solar System, $V_{\mathrm{eff}}$ has a minimum at $\phi = 0$, implying $A(\phi)=1$ and no direct coupling between DE and DM. This is an example of screening, allowing to suppress the fifth force inside the Solar System and stay in compliance with the existing tests of GR~\cite{Hinterbichler:2011ca}. 

Under the QSA, the scalar field always remains at the minimum of the effective potential. This condition leads to a useful expression for $\phi$ at densities below the point of  \emph{spontaneous symmetry breaking} (SSB),
	\begin{equation}
		\phi = \frac{\mu}{\sqrt{\lambda}} \sqrt{1 - \left( {a_{SSB}}/{a} \right)^3 },
		\label{eq:phi qsa}
	\end{equation}
where 	
	\begin{equation}
		a_{SSB}^3 = \frac{\rho_{\mathrm{dm}}^{(0)}}{A_0 \mu^2 M^2}.
		\label{eq:a_ssb}
	\end{equation} 
is the scale factor at the SSB. 
The three parameters $(\mu, \lambda , M)$ appearing in the symmetron Lagrangian can be expressed in terms of three more phenomenologically transparent quantities: the redshift at symmetry breaking, $z_{SSB}$, the present value of the coupling strength, $\beta_0$, and the Compton wavelength, $\lambda_C$, which sets the range of the DM-DE interaction. The relations are given by~\cite{Llinares:2014zxa} (see Appendix \ref{app:parameters symmetron} for more details) 
	\begin{subequations}
		\label{eq:symmetron obs parameters}
		\begin{align}
			\mu &= \frac{1}{\lambda_C \sqrt{2 \left(1 - a_{SSB}^3\right)}}, \label{eq:mu symmetron} \\
			M &= \frac{1}{\beta_0} \sqrt{\left[\sqrt{1 + \left(2\beta_0^2 \rho_{\mathrm{dm}}^{(0)} / \mu^2 a_{SSB}^3 \right)} -1 \right]}, \label{eq:M} \\
			\lambda &= \frac{1}{2 \lambda_C^2 \beta_0^2 M^4}. \label{eq:lambda}
		\end{align}
	\end{subequations}
The effective mass parameter appearing in Eq.~\eqref{eq:deltaphi_qsacdm} can be written as~\cite{Hojjati:2015ojt}
	\begin{equation}
		m = \mu \sqrt{2 \left[1 - \left( {a_{SSB}}/{a} \right)^3\right] }. \label{eq:m_qsa}
	\end{equation}

\section{Testing validity of the QSA} 
\label{ss:validity qsa}

To test the validity of the QSA, we compare the numerical solutions of the exact equations describing the evolution of linear perturbations in the symmetron model to their solutions obtained under the QSA, considering the cases of the scalar field coupled to DM and to all matter.
We are especially interested in the evolution of perturbations after the point of spontaneous symmetry breaking, when the symmetron scalar field develops a vacuum expectation value and starts mediating a fifth force between matter particles.

We fixed the value of the coupling parameter at $\beta_0=1$ and used the set of equations provided in Appendices \ref{ss:cdm_only} and \ref{app:trans_frames} to solve for the evolution of cosmological perturbations for a wide range of the other two symmetron model parameters: $1\le z_{SSB} \le 500$ and $0.1 \rm{Mpc} \le \lambda_C \le 100 \rm{Mpc}$. For the all-matter coupled case, we evolved the equations in the Einstein frame, given by \eqref{eq:exact_all}, and transformed the solutions to the baryon frame using \eqref{eq:frame transformations}. We set the initial conditions deep in the radiation era, as detailed in Appendix \ref{app:initial conditions}. When working under the QSA, we evolve $\mathrm{\Lambda CDM}$ equations up to the time of SSB, and switch to Eqs.~\eqref{eq:qsa_cdmonly} and \eqref{eq:qsa_all2} after the that.

\begin{figure*}[tbph!]
\includegraphics[width=0.49\textwidth]{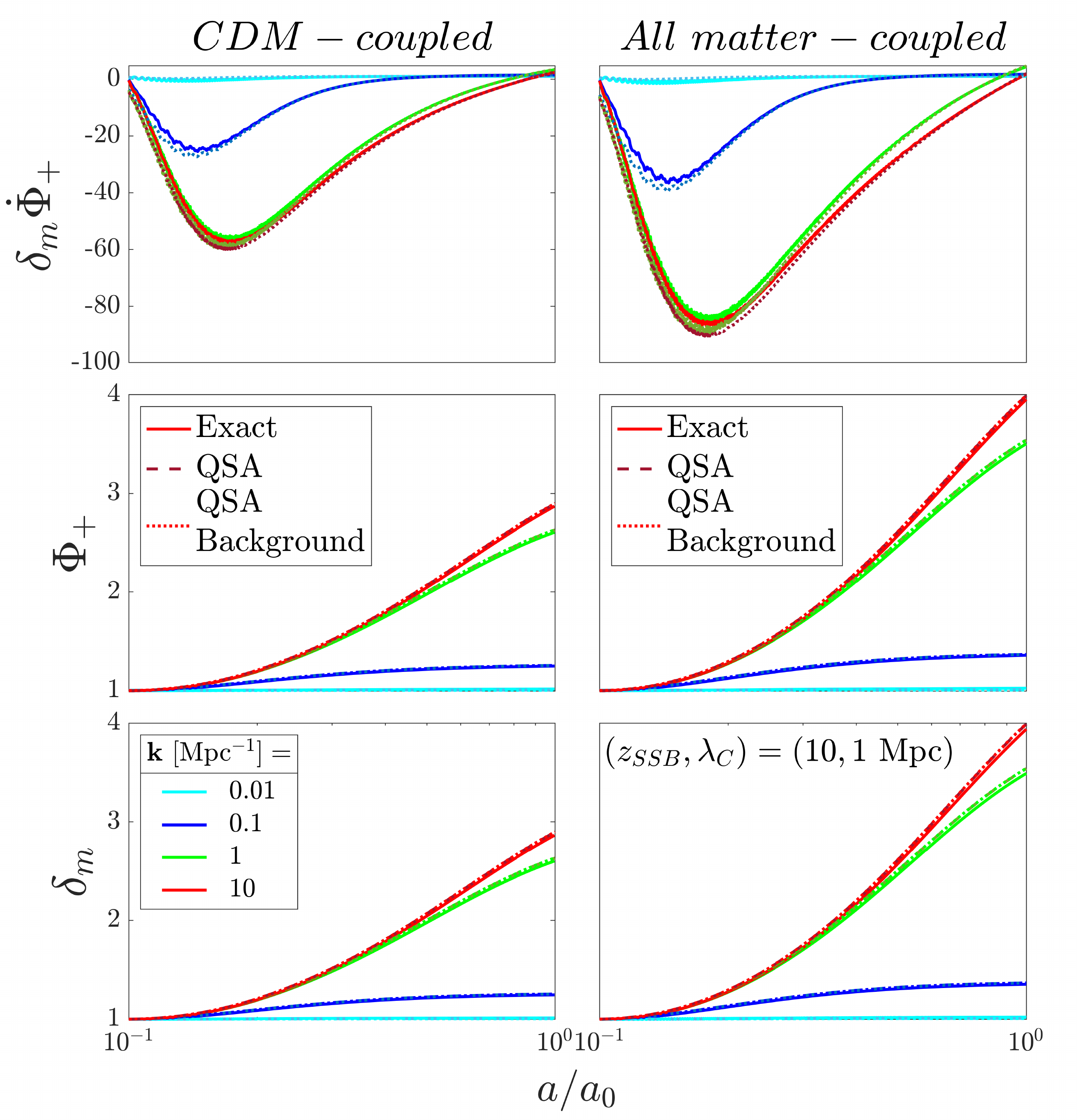}
\includegraphics[width=0.49\textwidth]{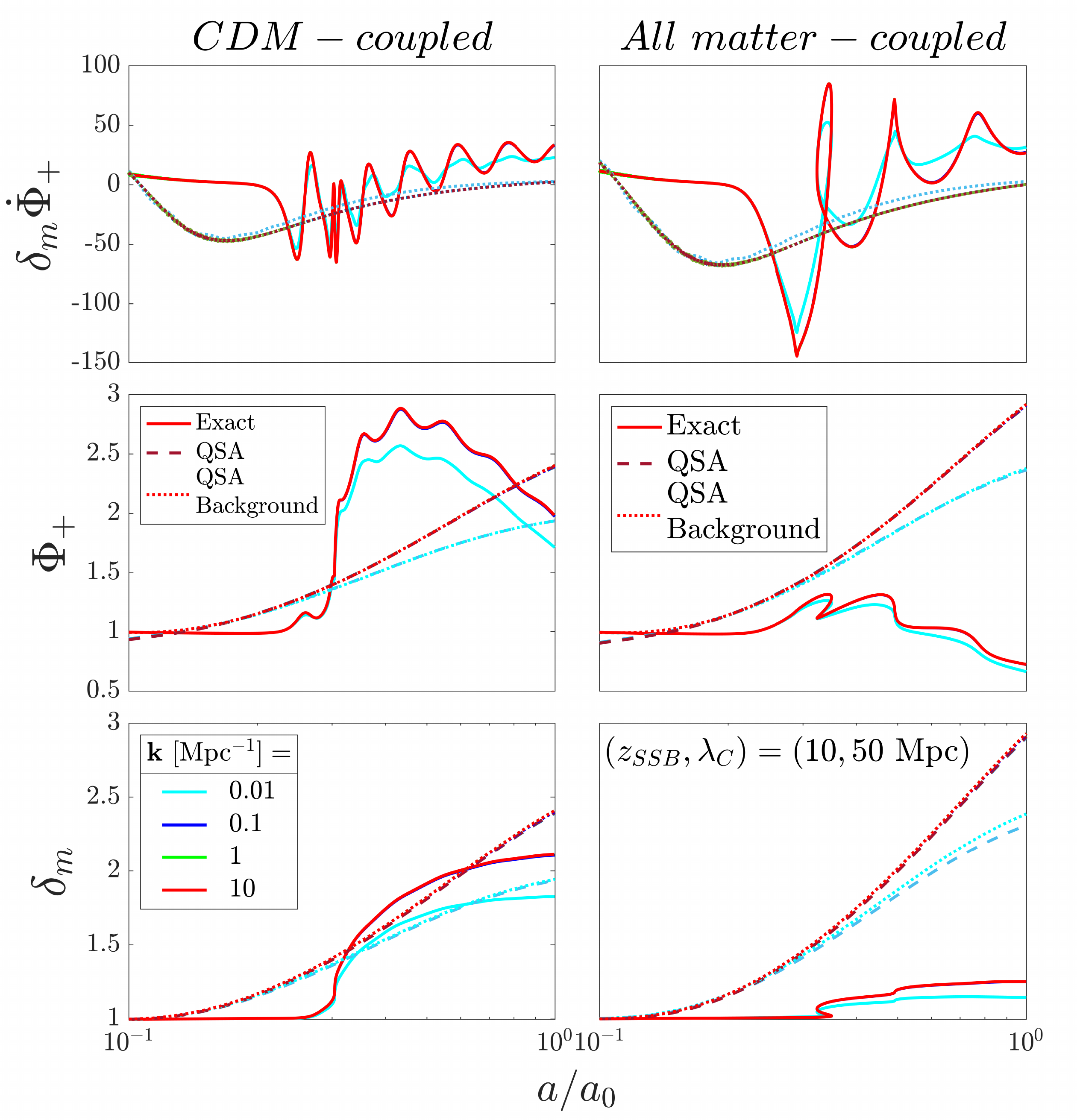}
\caption{\label{fig:perturbations_evolution} Relative departures from $\mathrm{\Lambda CDM}$ in the evolution of the matter density contrast $\delta_m$, the gravitational lensing potential $\Phi_{+}$ and the product of its time-derivative and $\delta_m$ (representative of the ISW-induced galaxy-CMB correlation) after the symmetry breaking at redshift $z = 10$ for four representative values of $k$. Shown are results for two symmetron models: $(z_{SSB},\lambda_C,\beta_0) = (10, 1~ \rm Mpc, 1~ M^{-1}_\mathrm{Pl})$ (left) and $(z_{SSB},\lambda_C,\beta_0) = (10, 50~ \rm Mpc, 1~ M^{-1}_\mathrm{Pl})$ (right), in the case of the DM-only and all-matter coupling. Different line types correspond to the exact solution (solid lines), the QSA (dashed lines), and the exact perturbations evolved on the QSA background (dotted lines). 
}
\end{figure*}

Fig.~\ref{fig:perturbations_evolution} compares the evolution of perturbations with the scale factor in two symmetron model that have the same $z_{SSB} = 10$ and $\beta_0=1$, but two different values of $\lambda_C$, for $k=0.01$, $0.1$, $1$, and $10$. We plot the matter density contrast, $\delta_m = (\rho_b\delta_b+\rho_{dm} \delta_{dm})/(\rho_b+\rho_{dm})$, the lensing potential $\Phi_+ = (\Phi+\Psi)/2$, and the product $\delta_m \dot{\Phi}_+$ that represents the strength of the galaxy-CMB correlation induced by the Integrated Sachs-Wolfe (ISW) effect. All the results are shown relative to the corresponding solution in the $\mathrm{\Lambda CDM}$ model, and we consider both the DM-only and the all-matter coupled cases. In the left block of panels, where the range of interaction between the matter and DE is smaller ($\lambda_C=1 \rm Mpc$), there is a good agreement between the exact and the QSA solutions for all the values of $k$-modes. We also see that, as expected, the modifications relative to $\mathrm{\Lambda CDM}$ are stronger in the case of the all-matter coupling. On the other hand, in the right block of panels, corresponding to $\lambda_C = 50 \rm Mpc$, we clearly see that the QSA approximation breaks down.

To separate the effects of assuming the QSA for the evolution of the background from the QSA for the perturbations, we also evolved the exact equations for linear perturbations on the QSA background, and the QSA perturbations on the exact background. This test revealed that, when using the QSA for the background, it makes no difference whether one uses the exact or the QSA perturbations. On the other hand, when using the exact background, the QSA breaks down in the same way whether one uses the exact or the QSA equations for the perturbations. This is illustrated in Fig.~\ref{fig:perturbations_evolution} which includes the results for the combination of the QSA background and exact perturbations, along with the fully exact and fully QSA solutions.

\begin{figure}[!th]
	\centering
	\includegraphics[width=0.85\textwidth]{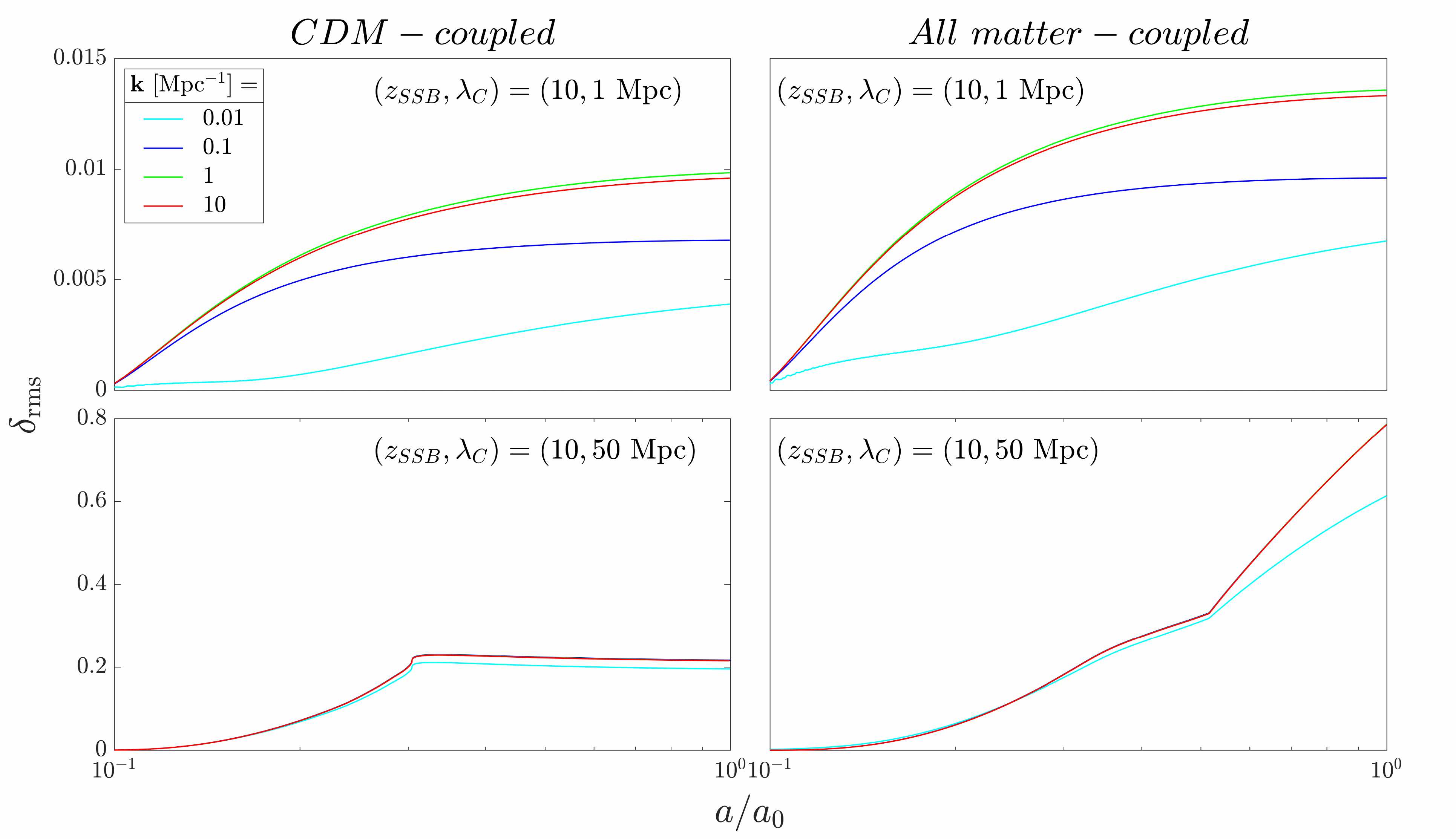}
	\caption{The rms fractional difference in the matter density contrast $\delta_m$ between the exact solution and the QSA after the symmetry breaking at redshift $z =10$ for four representative values of $k$. Shown are results for two symmetron models: $(z_{SSB},\lambda_C,\beta_0) = (10, 1~ \rm Mpc, 1~ M^{-1}_\mathrm{Pl})$, and $(z_{SSB},\lambda_C,\beta_0) = (10, 50~ \rm Mpc, 1~ M^{-1}_\mathrm{Pl})$, in the case of the DM-only coupling (left), and all-matter coupling (right).}
	\label{fig:rms}
\end{figure}

To assess the accuracy of the QSA, we calculated the root-mean-square ($\rm{rms}$) of the fractional difference in the matter density contrast $\delta_m$ between the exact solution and the QSA for two symmetron models depicted in Fig.~\ref{fig:perturbations_evolution}. The corresponding results are presented in Fig.~\ref{fig:rms}. The top two plots demonstrate the effectiveness of the QSA for $\lambda_C = 1 \rm Mpc$, with differences of less than 1\% observed for the DM-only coupled case and less than 1.5\% for the all-matter coupled case when comparing the exact solution with the QSA. Conversely, the bottom two plots, corresponding to $\lambda_C = 50 \rm Mpc$, reveal a significant discrepancy. The discrepancy reaches 20\% for the CDM-coupled case and extends to 80\% for the all-matter coupled case, indicating a failure of the QSA in these scenarios.

\begin{figure}[!th]
		\centering
\includegraphics[width=0.49\textwidth]{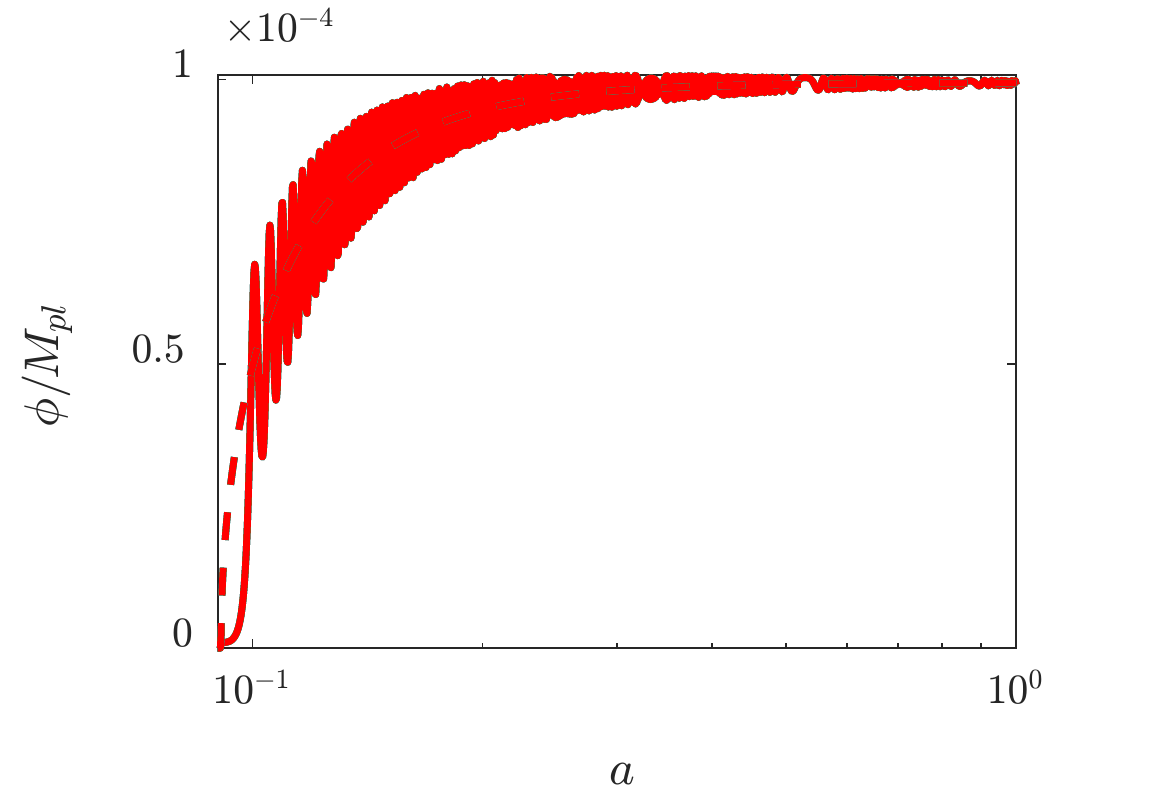}
\includegraphics[width=0.49\textwidth]{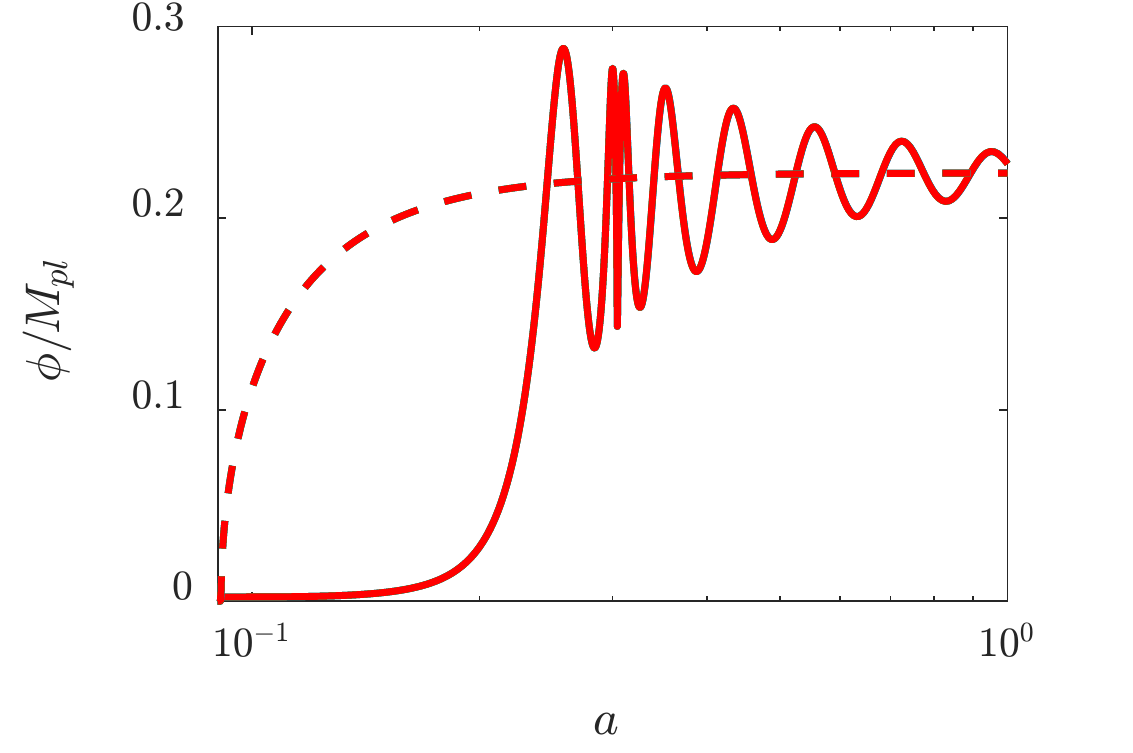}
\caption{Evolution of the background scalar field for the two cases shown in Fig.~\ref{fig:perturbations_evolution}, with $(z_{SSB},\lambda_C,\beta_0) = (10, 1~ \rm Mpc, 1~ M^{-1}_\mathrm{Pl})$ (left) and $(z_{SSB},\lambda_C,\beta_0) = (10, 50~ \rm Mpc, 1~ M^{-1}_\mathrm{Pl})$ (right), with the solid lines representing the exact solution and dash lines for QSA.}
\label{fig:evolution_of_phi}
\end{figure}

Examining the solutions for the two models shown in Fig.~\ref{fig:perturbations_evolution}, as well as our numerical results for other values of symmetron parameters, we find that the agreement between the exact and the QSA cases depends primarily on the dynamics of the background value of the scalar field after the point of symmetry breaking. Fig.~\ref{fig:evolution_of_phi} clearly shows the difference between the two cases. A larger Compton wavelength parameter, $\lambda_C \sim m^{-1}$, causes a longer delay in response of $\phi$ to the change in the shape of the effective potential at the SSB. It also reduces the frequency of oscillations of the background field around the average value predicted by the QSA solution, and allows for more freedom, {\it i.e.} a larger amplitude, for the scalar field to oscillate around that average due to a flatter shape of the effective potential at the minimum, since $V''_{\rm eff} = m^2$. 

We note that Eq.~\eqref{eq:thetadotdm_cdm} suggests that QSA may not be a good approximation when $\beta \dot\phi \sim \mathcal{H}$. However, our analysis shows that, for the symmetron model, the Hubble friction is always dominant in this equation even for the large values of $\lambda_C$.

\begin{figure}[!th]
	\centering
	\includegraphics[width=0.7\textwidth]{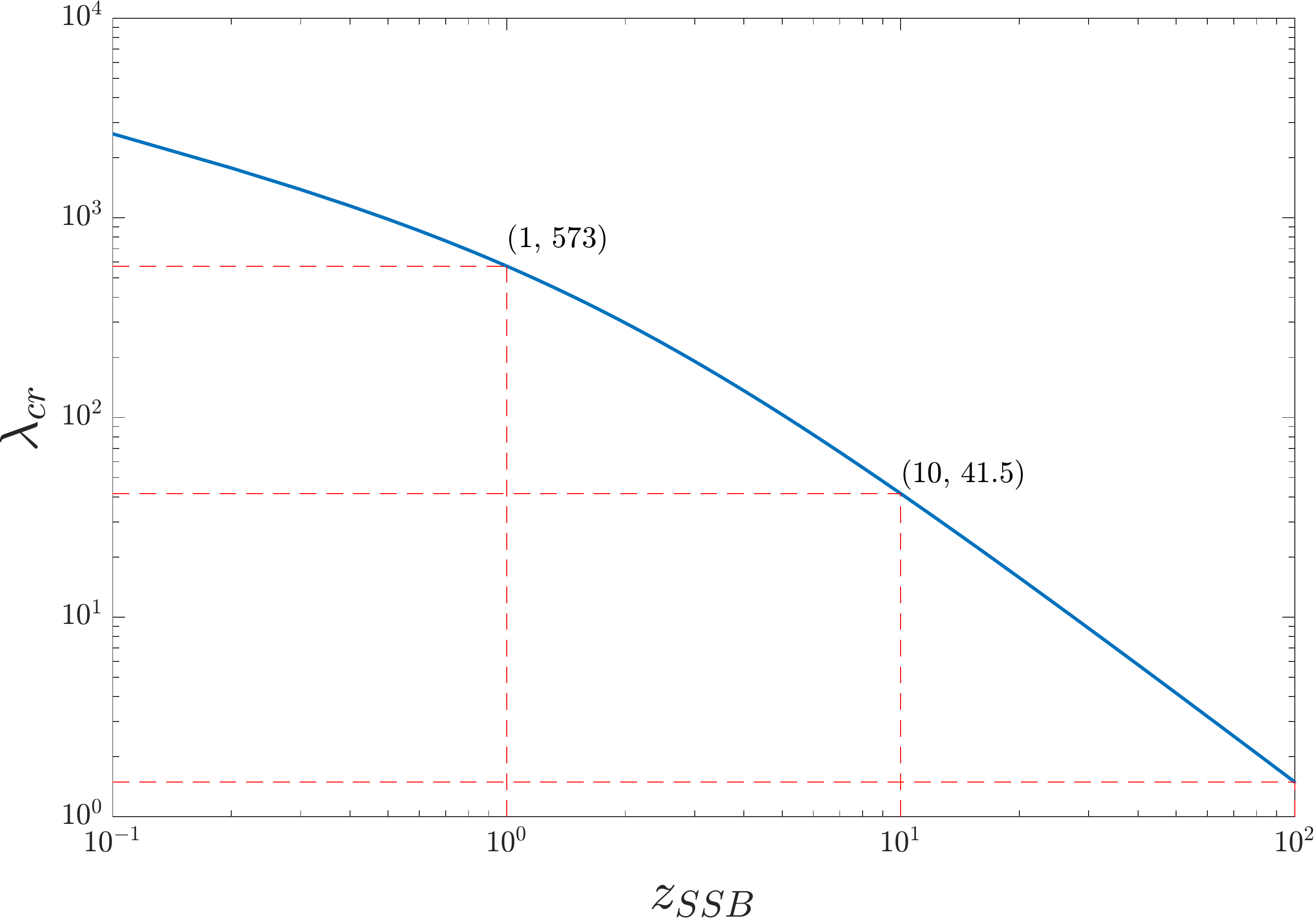}
	\caption{The values of the critical Compton wavelength $\lambda_{cr} [\rm{Mpc}]$ at which the QSA breaks down vs. the symmetry breaking redshift}
	\label{fig:lambdacrit}
\end{figure}
	
Our numerical results show that the QSA holds when $1 \le z_{SSB} \le 10$ and $\lambda_C<10$ Mpc, with the best consistency for $\lambda_C = 1$ Mpc. This range of validity of the QSA also follows from a simple analytical consideration presented below.

Prior to specializing to the symmetron example, it helps to note that the QSA assumes that the background scalar field follows the minimum of the effective potential $V_{\rm{eff}}(\phi)$, which includes a term proportional to the decreasing background matter density. As a result, the minimum of $V_{\rm{eff}}$ is continuously shifting and the QSA assumes that the scalar field manages to keep up with the shift. In addition, the scalar field oscillates around the minimum, which is not described by the QSA. However, in most cases of cosmological relevance, these oscillations do not affect the validity of the QSA when it comes to evaluating observable quantities, as the oscillations tend to average out to zero over cosmological timescales. When the scalar field evolution is dominated by the Hubble friction, it is unable to keep up with the shift in the minimum, and its average value becomes inconsistent with the prediction of the QSA. Hence, the criterion for the validity of the QSA is set by the balance between the time-scale $m^{-1}$, set by the curvature of the effective potential $V''_{\mathrm{eff}}$, and the Hubble scale $H^{-1}$. These considerations are generally applicable to all scalar-tensor theories of generalized Brans-Dicke type~\cite{Hinterbichler:2011ca, Khoury:2013yya}.

Let us now consider $V_{\rm{eff}}(\phi)$ of the symmetron model near $\phi = 0$ right after the point of symmetry breaking. Since the value of the scalar field was close to zero at that point, we can approximately write $V_{\rm{eff}}(\phi)$ as
\begin{equation}
		\label{eq:v_eff_approx}
		V_{\mathrm{eff}} \sim 
		V_0 + \frac{\rho_{\mathrm{dm}}^{(0)}}{A_0 a^3}
		+ \frac{1}{2}\phi^2 \left( - \mu^2 + \frac{\rho_{\mathrm{dm}}^{(0)}}{A_0 M^2 a^3} \right) \sim V^0_{\mathrm{eff}} - \frac{1}{2} \mu^2 \phi^2,
\end{equation}
where $V^0_{\mathrm{eff}} = V_0 + {\rho_{\mathrm{dm}}^{(0)}}/(A_0 a^3)$, and we have omitted the $\lambda \phi^4$ term since it is small compared to the $\phi^2$ terms, and also ignored the matter density term, $\rho_{\mathrm{dm}}^{(0)}/(A_0 M^2 a^3)$, as it decreases rapidly with $a$ and quickly becomes much smaller than $\mu^2$. The equation of motion for the scalar field right after the SSB can be written as
	\begin{equation}
		\label{eq:deeom_approx}
		\ddot{\phi} + 2 \mathcal{H} \dot{\phi} - \mu^2 a^2 \phi = 0.
	\end{equation}
We consider two regimes for the transition from a zero expectation value of $\phi$ to a none-zero value at the SSB: a fast transition, where $\ddot{\phi} \gg \mathcal{H} \dot{\phi}$ and the Hubble friction can be neglected, and a slow transition dominated by the Hubble friction, with $\ddot{\phi} \ll \mathcal{H} \dot{\phi}$. We can estimate the time-scale of the transition in the slow and the fast cases, $T_s$ and $T_f$, as
\begin{equation}
		\label{eq:transition time}
		T_s \sim \frac{\phi}{\dot{\phi}} \sim \frac{2 \mathcal{H}}{\mu^2 a^2}, \qquad T^2_f \sim \frac{\phi}{\ddot{\phi}} ~ \sim \frac{1}{\mu^2 a^2}.
	\end{equation} 
We expect the QSA to break down when the slow and fast transition time-scales are of the same order, {\it i.e.} $T_s \sim T_f$ or $2\mathcal{H}_{SSB} \sim \mu a_{SSB}$. From this condition and Eq.~\eqref{eq:mu symmetron}, which relates $\mu$ to $\lambda_C$ and $a_{SSB}$, we can derive an expression for the critical value of the Compton length above which the QSA fails:
\begin{equation}
		\label{eq:lambdacr}
		\lambda_{cr} \sim \frac{1}{H_0 \sqrt{8 [(1+z_{SSB})^3 - 1]}},
\end{equation}
where $H_0$ is the Hubble constant, and we used $\mathcal{H}_{SSB} \sim H_0 \sqrt{1 + z_{SSB}}$ assuming the symmetry breaking occurs during matter domination. The plot of $\lambda_{cr}$ vs $z_{SSB}$ in Fig.~\ref{fig:lambdacrit} shows that the analytically determined range of validity of the QSA, $\lambda_C < \lambda_{cr}$, is in good agreement with our numerically obtained results, such as the two examples shown in Fig.~\ref{fig:perturbations_evolution}.

\section{Summary} 
\label{s:discussion}

Scalar-tensor theories of gravity provide a well motivated framework to study the universe beyond the $\mathrm{\Lambda CDM}$ model. Among such theories, the symmetron model is a compelling example with a screening mechanism and a renormalizable potential, making it an excellent prototype for late dark energy. 

Although many studies of observational constraints on scalar-tensor theories assumed the QSA to be valid in the late universe, there is no guarantee that it holds for all ranges of the parameters in specific modified gravity model. In this work, we focused on the symmetron model to examine in detail the range of validity of the QSA. We found that it is the evolution of the background scalar field that is most important. Namely, whether its dynamics is determined by the Hubble friction or the scalar field potential. In the friction dominated case, $m \ll H$, the field is unable to keep up with the shifting minimum of the effective potential, and the QSA is broken. In the opposite limit, the QSA works. We demonstrated this numerically by comparing the exact solutions for the background and perturbations in the symmetron model with the QSA prediction. We also derived an expression (\ref{eq:lambdacr}) for the critical Compton wavelength, given the redshift of SSB, above which the QSA breaks down. Our arguments can be applied to derive similar criteria in other scalar-tensor theories of generalized Brans-Dicke type.
	
Our analysis paves the way for using MGCAMB and MGCLASS, that assume the QSA, for deriving observational constraints on the parameters of the symmetron model and other scalar-tensor theories.

\appendix
	
\section{Background and perturbations in scalar-tensor theories} \label{app:perturbations set}
	
	Here we present the complete set of equations that we use to evolve the background and perturbations for the exact case.
	
	\subsection{CDM-only coupled case} \label{ss:cdm_only}
	Varying Eq.~\eqref{eq:action} with respect to $g_{\mu\nu}$ yields the Einstein's field equations,
	\begin{equation}
		\label{eq:Modified Einstein}
		G_{\mu\nu} = 8\pi G \left( T^\mathrm{SM}_{\mu\nu} + A^2 (\phi) \tilde{T}^{\mathrm{dm}}_{\mu\nu} + T^{\mathrm{\phi}}_{{\mu\nu}} \right),
	\end{equation}
	where $T^{\mathrm{SM}}_{\mu\nu}$ and $\tilde{T}^{\mathrm{dm}}_{\mu\nu}$ are the stress-energy tensors of normal matter and DM in baryon and DM frames, respectively, given by
	\begin{subequations}
		\begin{align}
		T^{\mathrm{SM}}_{\mu\nu} &= \frac{-2}{\sqrt{-g}} \frac{\delta \mathcal{L}_{\mathrm{SM}}}{\delta g^{\mu\nu}}, \label{eq:matter energy tensor} \\
		\tilde{T}^{\mathrm{dm}}_{\mu\nu} &= \frac{-2}{\sqrt{-\tilde{g}}} \frac{\delta \mathcal{L}_{\mathrm{dm}}}{\delta \tilde{g}^{\mu\nu}} = A^{-2} (\phi) T^{\mathrm{dm}}_{\mu\nu}, \label{eq:dm energy tensor}
		\end{align}
	\end{subequations}
	with Eq.~\eqref{eq:dm energy tensor} relating the DM stress-energy tensors in two frames. The scalar field stress-energy tensor is
	\begin{equation}
		\label{eq:scalar field tensor}
		T^{\phi}_{\mu\nu} = \partial_{\mu} \phi \, \partial_{\nu} \phi - g_{\mu\nu} \left( \frac{1}{2} \partial^{\alpha}\phi \, \partial_{\alpha} \phi + V(\phi) \right).
	\end{equation}
	Varying the action with respect to $\phi$ yields the scalar field equation of motion (EoM),
	\begin{equation}
		\label{eq:dark energy eom}
		\square{\phi} = V,_{\phi} - A^3(\phi) A, _{\phi} \, \tilde{T}_{\mathrm{dm}} \, ,
	\end{equation}
	where $\tilde{T}_{\mathrm{dm}}$ is the trace of $\tilde{T}^{\mathrm{dm}}_{\mu\nu}$ that transforms between two frames through
	\begin{equation}
		\label{eq:stress trans}
		\tilde{T}_{\mathrm{dm}} = A^{-4}(\phi) T_{\mathrm{dm}}.
	\end{equation} 
	We will assume that, in the DM frame, the DM acts as dust, {\it i.e.} a perfect fluid with no pressure.
	Since we are writing our equations in the baryon frame, which is the co-moving frame for normal matter, the stress-energy tensor of baryons, photons, and neutrinos is conserved,
	\begin{equation}
		\label{eq:m stress conserv}
		\nabla_{\mu} T^{\mu\nu}_{\mathrm{SM}} = 0.
	\end{equation}
	The \emph{Bianchi Identity} for the Einstein tensor $G_{\mu\nu}$ implies that the sum of the scalar field and DM stress-energies should be conserved, namely,
	\begin{equation}
		\label{eq:field and dm conserv}
		\nabla_{\mu} T^{\mu\nu}_{\mathrm{\phi}} = -\nabla_{\mu} T^{\mu\nu}_{\mathrm{dm}} \, ,
	\end{equation}
	which, after using Eqs.~\eqref{eq:scalar field tensor} and \eqref{eq:dark energy eom}, gives
	\begin{equation}
		\label{eq:dark matter conservation}
		\nabla_{\mu} T^{\mu}_{\nu (\mathrm{dm})} = \frac{A,_{\phi}}{A} T_{\mathrm{dm}} \, \partial_{\nu} \phi.
	\end{equation}
	The effect of the fifth force on DM particles enters through the right hand side of Eq.~\eqref{eq:dark matter conservation} and vanishes when $A,_{\phi}=0$.
	The background expansion is given by the Friedmann equation,
	\begin{equation}
		\label{eq:friedmann background}
		\mathcal{H}^2 = \frac{8\pi G a^2}{3} 
		\left(
		\frac{\dot{\phi}^2}{2 a^2} + V(\phi) + \rho_{\mathrm{dm}} + \rho_{b} + \rho_{\gamma} + \rho_{\nu} 
		\right).
	\end{equation}
	The background evolution of the scalar field is given by
	\begin{equation}
		\label{eq:de eom back}
		\ddot{\phi} + 2 \mathcal{H} \dot{\phi} = -a^2 \left( V,_{\phi} + \frac{A,_{\phi}}{A} \rho_{\mathrm{dm}} \right),
	\end{equation}
	and the DM density evolves according to
	\begin{equation}
		\label{eq:dm eom back}
		\dot{\rho}_{\mathrm{dm}} + 3 \mathcal{H} \rho_{\mathrm{dm}} = \frac{A,_{\phi}}{A} \rho_{\mathrm{dm}} \dot{\phi} \, .
	\end{equation}
	For numerical implementation in codes such as CAMB~\cite{Lewis:1999bs}, it helps to write the above equations as 
	\begin{subequations}
		\label{eq:background set}
		\begin{align}
			\mathcal{H}^2 &= \frac{a^2}{3M^2_{\mathrm{Pl}}} \left( \frac{\dot{\phi}^2}{2 a^2} + V(\phi) \right) + \mathcal{H}^2_0 a^2 \left( \Omega_{\mathrm{dm}} + \frac{\Omega^{(0)}_\mathrm{b}}{a^3} +\frac{\Omega^{(0)}_{\gamma}}{a^4} +\frac{\Omega^{(0)}_{\nu}}{a^4} \right) , \label{eq:Friedmann Mpc} 
			\\
			\ddot{\phi} &= -2 \mathcal{H} \dot{\phi} -a^2 \left( V,_{\phi} + 3\mathcal{H}^2_0 \frac{A,_{\phi}}{A} \Omega_{\mathrm{dm}} \right),  \label{eq:DE background Mpc} 
			\\
			\dot{\Omega}_{\mathrm{dm}} &= \left( \frac{A,_{\phi}}{A} \dot{\phi} - 3 \mathcal{H} \right) \Omega_{\mathrm{dm}}, \label{eq:DM background Mpc}		
		\end{align}
	\end{subequations}
	where $\Omega^{(0)}_{\mathrm{i}}$ is the density parameter for an individual component at the present epoch \cite{Copeland:2006wr}, and $\Omega_{\mathrm{dm}} \equiv \rho_{\mathrm{dm}}/(3 M^2_{\mathrm{Pl}} \mathcal{H}^2_0)$. Here, $\phi$ has units of $M_\mathrm{pl}$ and the conformal time is in units of megaparsec [Mpc]. Eqs.~\eqref{eq:background set} form a complete set needed to solve for the background evolution.
	
	Eqs.~\eqref{eq:Modified Einstein}, \eqref{eq:metric per}, and \eqref{eq:de parts} imply four equations for the linear metric perturbations in Fourier space, but one only needs two out of four Einstein equations given by
	\begin{subequations}
		\label{eq:einstein eq per}
		\begin{align}
			\begin{split}
				k^2 \Phi + 3\mathcal{H} (\dot{\Phi} + \mathcal{H} \Psi) &= -4 \pi Ga^2 
				\Biggl( 
				\rho_b \delta_{b} + \rho_{\mathrm{dm}} \delta_{\mathrm{dm}} + \rho_{\gamma} \delta_{\gamma} + \rho_{\nu} \delta_{\nu} \\
				&\quad+ \frac{\dot{\phi} \delta{\dot{\phi}}}{a^2} + V,_{\phi} \delta{\phi} - \Psi \frac{\dot{\phi}^2}{a^2} 
				\Biggr), 
				\label{eq:einstein 00}
			\end{split} \\
			k^2 (\Phi - \Psi) &= 16 \pi G a^2 ( \rho_{\gamma} \sigma_{\gamma} + \rho_{\nu} \sigma_{\nu}), \label{eq:einstein ij}
		\end{align}
	\end{subequations}
	with the choice being a matter of convenience. We select \eqref{eq:einstein 00} and the derivative of \eqref{eq:einstein ij}, since both $\dot{\Psi}$ and $\dot{\Phi}$ appear in Eq.~\eqref{eq:de eom per}.
	Using Eqs.~\eqref{eq:metric per} and \eqref{eq:de parts} in \eqref{eq:dark energy eom}, yields the EoM for scalar field perturbations:
	\begin{equation}
		\label{eq:de eom per}
		\begin{split}
			\delta \ddot{\phi} + 2 \mathcal{H} \delta \dot{\phi} + k^2 \delta \phi =& 
			-a^2 \Biggl\{
			2 \Psi \left( V,_{\phi} + \frac{A,_{\phi}}{A} \rho_{\mathrm{dm}} \right) +
			\left[ V,_{\phi \phi} + \left( \frac{A,_{\phi \phi}}{A} - (\frac{A,_{\phi}}{A})^2 \right) \rho_{\mathrm{dm}} \right] \delta \phi \\
			&+ \frac{A,_{\phi}}{A} \rho_{\mathrm{dm}} \delta_{\mathrm{dm}}
			\Biggr\}
			+ \dot{\phi} \left( \dot{\Psi} + 3\dot{\Phi} \right).
		\end{split}
	\end{equation}
	Perturbing Eq.~\eqref{eq:dark matter conservation} gives the continuity and Euler equations for the DM perturbations:
	\begin{subequations}
		\label{eq:dm per}
		\begin{align}
			\dot{\delta}_{\mathrm{dm}} &=
			- \theta_{\mathrm{dm}} + 3\dot{\Phi} + \frac{A,_{\phi}}{A} \delta \dot{\phi} 
			+ \left[ \frac{A,_{\phi \phi}}{A} - (\frac{A,_{\phi}}{A})^2 \right] \dot{\phi} \delta \phi,
			\label{eq:dm energy per}
			\\
			\dot{\theta}_{\mathrm{dm}} &= - \left( \mathcal{H}
			+ \frac{A,_{\phi}}{A} \dot{\phi} \right) \theta_{\mathrm{dm}}  + k^2 \left( \Psi + \frac{A,_{\phi}}{A} \delta \phi \right).
		\end{align}
	\end{subequations}
	For the baryons, the conservation equations are the same as in $\Lambda$CDM, since they do not couple to the scalar field:
	\begin{subequations}
		\label{eq:baryons per}
		\begin{align}
			\dot{\delta}_b + \theta_b &= 3 \dot{\Phi}, \label{eq:baryons energy per}
			\\
			\dot{\theta}_b + \mathcal{H} \theta_b &= k^2 \Psi
			+ c_s^2 k^2 \delta_b + \frac{\dot{\tau}_c}{R} 
			\left( \theta_{\gamma} - \theta_b \right),
		\end{align}
	\end{subequations}
	The quantity $c_s$ is the baryon sound speed given by
	\begin{equation}
		\label{eq:sound speed}
		c_s^2 = \frac{dP}{d \rho} = \frac{1}{3(1+R)},
	\end{equation}
	where $R = 3\rho_b/4\rho_{\gamma}$ is the baryon-photon density ratio. $\dot{\tau}_c = a n_e \sigma_T$ is the differential optical depth, $n_e$ is the free electron density and $\sigma_T$ is the Thomson scattering cross section. The Thomson opacity is important before and around the epoch of recombination, when the photons and baryons are tightly coupled   \cite{Ma:1995ey}. In \cite{Hu:1994uz}, Hu and Sugiyama provided a fitting formula for $\dot{\tau}_c$
	\begin{equation}
		\dot{\tau}_c^{-1} = 4.3 \times 10^4 \left( 1 - \frac{Y_p}{2} \right)^{-1} 
		\left( \Omega_b h^2 \right)^{-1} \left( 1 + z \right)^{-2} \mathrm{Mpc},
	\end{equation}
	which can be used for quick estimates once the ionization history is provided. In the above, $Y_p \approx 0.23$ is the primordial helium mass fraction and $h$ is the dimensionless Hubble parameter.
	
	The photon and neutrino phase-space distributions evolve according to the perturbed Boltzmann equations and can be expressed in terms of perturbations in their corresponding black-body temperatures. 
	Expanding the temperature anisotropies in Legendre  multiple moments gives an infinite set of coupled differential equations known as the \emph{Boltzmann hierarchy} \cite{Seljak:1996is}. The hierarchy can be truncated at some maximum multiple moment $l_{max}$ chosen according to the desired numerical accuracy. For our purposes, it will be sufficient to work with $l_{max} = 3$.
	
	A full list for evolution of perturbations is given by
	\begin{subequations}
		\renewcommand{\theequation}{\theparentequation.\arabic{equation}}
		\label{eq:perturbations set}
		\begin{align}
			\dot{\Phi} &= -\mathcal{H} \Psi + \frac{3 \mathcal{H}_0^2 a^2}{2k^2} \Biggl[
			\frac{\Omega_b^{(0)}}{a^3} \theta_b + \Omega_{\mathrm{dm}} \theta_{\mathrm{dm}} + \frac{4}{3} \Biggl( \frac{\Omega_{\nu}^{(0)}}{a^4} \theta_{\nu} + \frac{\Omega_{\gamma}^{(0)}}{a^4} \theta_{\gamma} \Biggr) \Biggr] + \frac{1}{2} \dot{\phi} \delta \phi
			, \label{eq:Phidot Mpl}
			\\
			\dot{\delta}_{\gamma} &= -\frac{4}{3} \theta_{\gamma} + 4 \dot{\Phi}, \label{eq:deltadot gamma Mpl}
			\\
			\dot{\theta}_{\gamma} &= k^2 \left( \frac{1}{4} \delta_{\gamma} - \sigma_{\gamma} \right) + \dot{\tau}_c \left( \theta_b - \theta_{\gamma} \right) + k^2 \Psi,	\label{eq:thetadot gamma Mpl}
			\\
			\dot{\delta}_{\mathrm{dm}} &= -\theta_{\mathrm{dm}} +
			3\dot{\Phi} + \frac{A,_{\phi}}{A} \delta \dot{\phi} 
			+ \left[ \frac{A,_{\phi \phi}}{A} - (\frac{A,_{\phi}}{A})^2 \right] \dot{\phi} \delta \phi,
			\label{eq:deltadot dm Mpl}	
			\\
			\dot{\theta}_{\mathrm{dm}}  &= - \left( \mathcal{H}
			+ \frac{A,_{\phi}}{A} \dot{\phi} \right) \theta_{\mathrm{dm}} + k^2 \left( \Psi + \frac{A,_{\phi}}{A} \delta \phi \right), \label{eq: thetadot dm}
			\\
			\dot{\delta}_b &= - \theta_b + 3 \dot{\Phi}, \label{eq:deltadot b}
			\\
			\dot{\theta}_b &= - \mathcal{H} \theta_b + k^2 \Psi
			+ c_s^2 k^2 \delta_b + \frac{\dot{\tau}_c}{R} 
			\left( \theta_{\gamma} - \theta_b \right),
			\label{eq:thetadot b}
			\\
			\dot{\delta}_{\nu} &= -\frac{4}{3} \theta_{\nu} + 4 \dot{\Phi}, \label{eq:deltadot nu}
			\\
			\dot{\theta}_{\nu} &= k^2 \left( \frac{1}{4} \delta_{\nu} - \sigma_{\nu} \right) + k^2 \Psi,
			\label{eq:thetadot nu}
			\\
			\dot{\sigma}_{\nu} &= \frac{4}{15} \theta_{\nu} - \frac{3}{10} k F_{\nu3}, \label{eq:sigmadot nu}
			\\
			\dot{F}_{\nu3} &= \frac{6}{7}k \sigma_{\nu}, \label{eq:Fdot nu3}
			\\
			\dot{\sigma}_{\gamma} &= \frac{4}{15} \theta_{\gamma} - \frac{3}{10} k F_{\gamma3} - \frac{9}{10} \dot{\tau}_c \sigma_{\gamma}, \label{eq:sigmadot gamma}
			\\
			\dot{F}_{\gamma3} &= \frac{6}{7}k \sigma_{\gamma} - \dot{\tau}_c F_{\gamma3} , \label{eq:Fdot gamma3}
			\\
			\dot{\Psi} &= \dot{\Phi} - \frac{6 \mathcal{H}_0^2 a^2}{k^2} \Biggl[ \frac{\Omega_{\nu}^{(0)}}{a^4}
			\left( \dot{\sigma}_{\nu} - 2\mathcal{H} \sigma_{\nu} \right) + \frac{\Omega_{\gamma}^{(0)}}{a^4}
			\left( \dot{\sigma}_{\gamma} - 2\mathcal{H} \sigma_{\gamma} \right)
			\Biggr],
			\label{eq:Psidot Mpl}
			\\
			\begin{split}
				\delta \ddot{\phi} &= 
				-2 \mathcal{H} \delta \dot{\phi} - k^2 \delta \phi-a^2 \Biggl\{
				2 \Psi \left( V,_{\phi} + 3\mathcal{H}_0^2 \frac{A,_{\phi}}{A} \Omega_{\mathrm{dm}} \right) 
				\\
				&\quad+ \left[ V,_{\phi \phi} + 3\mathcal{H}_0^2 \left( \frac{A,_{\phi \phi}}{A} - (\frac{A,_{\phi}}{A})^2 \right) \Omega_{\mathrm{dm}} \right] \delta \phi  + 3\mathcal{H}_0^2 \frac{A,_{\phi}}{A} \Omega_{\mathrm{dm}} \delta_{\mathrm{dm}}
				\Biggr\}
				+ \dot{\phi} \left( \dot{\Psi} + 3\dot{\Phi} \right),
				\label{eq:dphiddot Mpl}
			\end{split}
		\end{align}
	\end{subequations}
	where following the truncation scheme provided by Ma \& Bertschinger in \cite{Ma:1995ey}, the equations for the photon and neutrino anisotropies can be written as \eqref{eq:sigmadot nu} to \eqref{eq:Fdot gamma3}. Here, $F_{\gamma l}$ and $F_{\nu l}$ denote the photon and neutrino multiple moments of order $l$, with $F_0 = \delta, F_1 = 4\theta/3k$, and $F_2 = 2 \sigma$. Note that the evolution of DM is modified via coupling to the scalar field as \eqref{eq:deltadot dm Mpl}, \eqref{eq: thetadot dm} while baryons are only coupled to photons in \eqref{eq:deltadot b}, \eqref{eq:thetadot b}.
	
	\subsection{The all-matter coupled case} \label{app:trans_frames}
	
	For the all-matter coupled case, we evolve the equations in the Einstein frame, and transform the solutions to the baryon frame.
	A conformal transformation transforms the Einstein frame metric $\tilde{g}_{\mu\nu}$ to baryon frame metric $g_{\mu\nu}$ via
	\begin{equation}
		g_{\mn} = A^2 (\tilde{\phi}) \tilde{g}_{\mn}. \label{eq:metrictrans_all}
	\end{equation}
	Hence, the line element transforms as
	\begin{equation}
		\label{eq:metric per jordan}
		ds^2 = a^2(\tau) \left[ -(1 + 2\Psi) d\tau^2 + (1-2\Phi) d \vec{x}^2 \right] = A^2(\tilde{\phi}) d\tilde{s}^2.
	\end{equation}
	Since $d\vec{x}$ is a comoving coordinate, it is not subject to change of scale in conformal transformation. This immediately gives two expressions relating the scale factor and conformal time between the baryon and Einstein frames as
	\begin{equation}
		\label{eq:a, tau jordan}
		a = A(\tilde{\phi}) \tilde{a}, \qquad d \tau = d\tilde{\tau},
	\end{equation}
	{\it i.e.} the conformal coordinate $\tau$ is the same in the two frames. The transformation between the cosmological variables in the two frames can be derived from requiring the action to remain invariant, {\it i.e.} $S = \tilde{S}$. Furthermore, we require the scalar field to have the same form of the kinetic and potential energy terms, giving the transformation for the scalar field and its potential~\cite{fujii_maeda_2003}:
	\begin{subequations}
		\label{eq:frame trans}
		\begin{align}
			\phi &= A^{-1} \tilde{\phi}, \label{eq:phi jordan} \\
			V(\phi) &= A^{-4} \tilde{V}(\tilde{\phi}). \label{eq:V jordan}
		\end{align}
	\end{subequations}
	Using Eq.~\eqref{eq:baryon energy tensor total} and the invariance of the action allows one to derive the transformation for the matter stress-energy:
	\begin{equation}
		\label{eq:stress tensor jordan}
		T_{\mu\nu} = A^{-2} \tilde{T}_{\mu\nu}, \qquad T = A^{-4} \tilde{T}.
	\end{equation}
	giving the transformations for the background pressure and energy density:
	\begin{equation}
		\label{eq:background energy pressure j}
		\rho = A^{-4} \tilde{\rho}, \qquad P = A^{-4} \tilde{P}.
	\end{equation}
	Since the stress-energy tensor of relativistic particles is traceless, the conservation equation for radiation density has the same form in the Einstein and baryon frames,
	\begin{equation}
		\label{eq:radiation eom jordan}
		\dot{\rho_r} + 4 \mathcal{H} \rho_r = 0.
	\end{equation}
	One can use this to derive a useful relation between the conformal Hubble parameter in the two frames:
	\begin{equation}
		\label{eq:conformal hubble jordan}
		\mathcal{H} = \tilde{\mathcal{H}} + \frac{A,_{\tilde{\phi}}}{A} \dot{\tilde{\phi}}.
	\end{equation}
	The transformations for the linear perturbations can be found by perturbing Eqs.~\eqref{eq:phi jordan} and \eqref{eq:background energy pressure j}, and from Eq.~\eqref{eq:metric per jordan}. The complete set of relations between the Einstein and baryon frames is given by	\begin{subequations}
		\renewcommand{\theequation}{\theparentequation.\arabic{equation}}
		\label{eq:frame transformations}
		\begin{align}
			\tilde{\beta} & \equiv \frac{A,_{\tilde{\phi}}}{A} = \frac{\beta}{A}, \qquad a = A\tilde{a}, \label{eq:beta and a j} \\
			\phi &= A^{-1} \tilde{\phi}, \qquad \dot{\phi} = A^{-1} \left( 1 - \tilde{\beta} \tilde{\phi} \right) \dot{\tilde{\phi}}, \label{eq:phi and phidot j} \\
			\delta \phi &= A^{-1} \left( 1 - \tilde{\beta} \tilde{\phi} \right) \delta \tilde{\phi}, \label{eq:delta phi j} \\
			\delta \dot{\phi} &= A^{-1} \left[ \left( 2 \tilde{\beta}^2 - \frac{A,_{\tilde{\phi} \tilde{\phi}}}{A} \right) \dot{\tilde{\phi}} \tilde{\phi} - 2 \tilde{\beta} \dot{\tilde{\phi}} \right] \delta \tilde{\phi} + A^{-1} \left( 1 - \tilde{\beta} \tilde{\phi} \right) \delta \dot{\tilde{\phi}}, \label{eq:delta phi dot j} \\
			\Phi &= \tilde{\Phi} - \tilde{\beta} \delta \tilde{\phi}, \qquad \Psi = \tilde{\Psi} + \tilde{\beta} \delta \tilde{\phi}, \label{eq:Phi and Psi j} \\
			\delta_{\gamma} &= \tilde{\delta}_{\gamma} -4 \tilde{\beta} \delta \tilde{\phi}, \qquad \delta_{\nu} = \tilde{\delta}_{\nu} - 4 \tilde{\beta} \delta \tilde{\phi}, \label{eq:delta radiation j} \\
			\delta_{\mathrm{dm}} &= \tilde{\delta}_{\mathrm{dm}} -4 \tilde{\beta} \delta \tilde{\phi}, \qquad \delta_{b} = \tilde{\delta}_{b} - 4 \tilde{\beta} \delta \tilde{\phi}, \label{eq:delta matter j} \\
			\theta_{\gamma} &= \tilde{\theta}_{\gamma}, \qquad \theta_{\nu} = \tilde{\theta}_{\nu}, \qquad
			\theta_{\mathrm{dm}} = \tilde{\theta}_{\mathrm{dm}}, \qquad \theta_{b} = \tilde{\theta}_{b}, \label{eq:theta j} \\
			\sigma_{\gamma} &= \tilde{\sigma}_{\gamma}, \qquad \sigma_{\nu} = \tilde{\sigma}_{\nu}. \label{eq:sigma j}
		\end{align}
	\end{subequations}
	The divergence of the fluid velocity $\theta$ and the shear stress $\sigma$ in k-space are defined as ~\cite{Ma:1995ey}
	\begin{equation}
		\label{eq:theta and sigma}
		(\rho + P) \theta \equiv i k^j \delta T^0_j, \qquad (\rho + P) \sigma \equiv -(\hat{k}_i . \hat{k}_j - \frac{1}{3} \delta_{ij}) \Sigma^i_j.
	\end{equation}
	It then follows from Eqs.~\eqref{eq:stress tensor jordan} and \eqref{eq:theta and sigma} that these two variables remain invariant under conformal transformations.
	Eqs.~\eqref{eq:frame trans}, \eqref{eq: scalarfield potential} directly relate the parameters in symmetron model between two frames by
	\begin{equation}
		\label{eq:symmetron parameters jordan}
		{V}_0 = A^{-4} \tilde{V_0}, \qquad {M} = A^{-1} \tilde{M}, \qquad \mu = A^{-1} \tilde{\mu}, \qquad {\lambda} = \tilde{\lambda}.
	\end{equation}

	The equations of motion when scalar field is coupled to all matter in the Einstein frame are given by
			\begin{subequations} \label{eq:exact_all}
				\renewcommand{\theequation}{\theparentequation.\arabic{equation}}
				\begin{align}
					&\mathcal{\tilde{H}}^2 = \frac{8\pi G \tilde{a}^2}{3} 
					\left(
					\frac{\dot{\tilde{\phi}}^2}{2 \tilde{a}^2} + \tilde{V}(\tilde{\phi}) + \tilde{\rho}_{\mathrm{dm}} + \tilde{\rho}_b + \tilde{\rho}_{\gamma} + \tilde{\rho}_{\nu} 
					\right), \label{eq:friedmann_all} \\
					&\ddot{\tilde{\phi}} + 2 \mathcal{\tilde{H}} \dot{\tilde{\phi}} = -\tilde{a}^2 \tilde{V}'_{\rm eff}, \label{eq:deeom_all} \\
					&\dot{\tilde{\rho}}_{\mathrm{dm}} + 3 \mathcal{\tilde{H}} \tilde{\rho}_{\mathrm{dm}} = {\tilde{\beta}} \dot{\tilde{\phi}} \tilde{\rho}_{\mathrm{dm}}, \label{eq:dmeom_all} \\
					&\dot{\tilde{\rho}}_b + 3 \mathcal{\tilde{H}} \tilde{\rho}_b = \dot{\tilde{\beta}} \tilde{\rho}_b, \label{eq:beom_all} \\
					%\begin{split}
						&k^2 \tilde{\Phi} + 3\mathcal{\tilde{H}} (\dot{\tilde{\Phi}} + \mathcal{\tilde{H}} \tilde{\Psi}) = -4 \pi G\tilde{a}^2 
						\Biggl( 
						{\sum\limits_{i} } \tilde{\rho}_{i} \tilde{\delta}_{i} + \tilde{V},_{\tilde{\phi}} \delta{\tilde{\phi}}
						+ \frac{\dot{\tilde{\phi}} \delta{\dot{\tilde{\phi}}}}{\tilde{a}^2} - \tilde{\Psi} \frac{\dot{\tilde{\phi}}^2}{\tilde{a}^2} 
						\Biggr), \label{eq:cphi_all} \\
					%\end{split} \\
					&k^2 (\tilde{\Phi} - \tilde{\Psi}) = 16 \pi G \tilde{a}^2 ( \tilde{\rho}_{\gamma} \tilde{\sigma}_{\gamma} + \tilde{\rho}_{\nu} \tilde{\sigma}_{\nu}), \label{eq:shear_all} \\
					&\dot{\tilde{\delta}}_b  = -\tilde{\theta}_b + 3 \dot{\tilde{\Phi}}+ \tilde{\beta} \delta \dot{\tilde{\phi}} 
					+ \dot{\tilde{\beta}} \delta \tilde{\phi},, \label{eq:deltadotb_all} \\
					%\begin{split}
						&\dot{\tilde{\theta}}_b = - \left( \mathcal{\tilde{H}}
						+ \tilde{\beta} \dot{\tilde{\phi}} \right) \tilde{\theta}_b + k^2 \left(\tilde{\Psi} + \tilde{\beta} \delta \tilde{\phi} \right)
						+ c_s^2 k^2 \tilde{\delta}_b + \frac{\dot{\tilde{\tau}}_c}{\tilde{R}} 
						\left( \tilde{\theta}_{\gamma} - \tilde{\theta}_b \right), \label{eq:thetadotb_all} \\
					%\end{split} \\
					&\dot{\tilde{\delta}}_{\mathrm{dm}} =
					- \tilde{\theta}_{\mathrm{dm}} + 3\dot{\tilde{\Phi}} + \tilde{\beta} \delta \dot{\tilde{\phi}} 
					+ \dot{\tilde{\beta}} \delta \tilde{\phi}, \label{eq:deltadotdm_all} \\
					&\dot{\tilde{\theta}}_{\mathrm{dm}} = - \left( \mathcal{\tilde{H}}
					+ \tilde{\beta} \dot{\tilde{\phi}} \right) \tilde{\theta}_{\mathrm{dm}}  + k^2 \left( \tilde{\Psi} + \tilde{\beta} \delta \tilde{\phi} \right), \label{eq:thetadotdm_all} \\
					%\begin{split}
						&\delta \ddot{\tilde{\phi}} + 2 \mathcal{\tilde{H}} \delta \dot{\tilde{\phi}} + k^2 \delta \tilde{\phi} = -\tilde{a}^2 
						\biggl\{
						2 \tilde{\Psi} \tilde{V}'_{\rm eff} +
						\tilde{V}''_{\rm eff} \delta \tilde{\phi}
						+ \tilde{\beta} \left( \tilde{\rho}_{\mathrm{dm}} \tilde{\delta}_{\mathrm{dm}} + \tilde{\rho}_{b} \tilde{\delta}_{b} \right) 
						\Biggr\}
						+ \dot{\tilde{\phi}} \left( \dot{\tilde{\Psi}} + 3\dot{\tilde{\Phi}} \right). \label{eq:ddeeom_all}
					%\end{split}
				\end{align}
			\end{subequations}
	where	
	\begin{align}
		\tilde{V}'_{\mathrm{eff}} &= \tilde{V},_{\tilde{\phi}} + \tilde{\beta} \left( \tilde{\rho}_{\mathrm{dm}} + \tilde{\rho}_{b} \right), \label{eq:vprimeeff_all}
	\end{align}
	is the derivative of effective potential with respect to $\tilde{\phi}$. The conversion of all the variables to the baryon frame is given by \eqref{eq:frame transformations}, \eqref{eq:symmetron parameters jordan}.

\subsection{Initial conditions} \label{app:initial conditions}
The system of coupled differential equations presented in earlier sections can be soled once the initial conditions are specified. We set them deep in the radiation era where the relevant Fourier modes are still outside the horizon. This satisfies the condition $k \tau \ll 1$. At the early time, the baryons and CDM have a negligible contribution to the energy density of the universe, so we have $\rho_{total} \approx \rho_{\nu} + \rho_{\gamma}$. The Hubble parameter is also inversely proportional to the conformal time and determined by $\mathcal{H} = \tau^{-1}$.
	We choose our initial conditions so that only the fastest-growing physical modes are present. This is appropriate for the perturbations that are created in early universe. We can then analytically find the time dependence of all metric and density perturbations by directly solving the differential equations similar to the ones in \cite{Ma:1995ey}. Using Eqs.~\eqref{eq:friedmann background}, \eqref{eq:perturbations set} for the isentropic perturbations in the conformal Newtonian gauge, we derive the following relations for the initial conditions:
	\begin{subequations}
		\renewcommand{\theequation}{\theparentequation.\arabic{equation}}
		\label{eq: initial conditions}
		\begin{align}
			\Psi &= \frac{20 C}{15 + 4 R_{\nu}},
			\label{eq: metric perini}
			\\
			\begin{split}
				\delta_{\gamma} &= - 2 \Psi \left( 1 - \frac{1}{6} \tau^2 \dot{\phi}^2 \right)
				- \frac{1}{3} \tau^2 \left( \dot{\phi} \delta \dot \phi + a^2 V,_{\phi} \delta \phi \right),
				\\
				\delta_{b} &= \frac{3}{4} \delta_{\gamma} = \frac{3}{4} \delta_{\nu}, \qquad
				\delta_{\mathrm{dm}} = \frac{3}{4} \delta_{\gamma} + \beta \delta \phi,
				\label{eq: deltaini}
			\end{split}
			\\
			\Phi &= \Psi \left( 1 + \frac{4}{5} R_{\nu} \right) + \frac{1}{5} R_{\nu} \delta_{\nu},
			\label{eq:Phi perini}
			\\
			\theta_{\gamma} &= \theta_{\nu} = \theta_b = \frac{1}{2} \left( k^2 \tau \right)\Psi, \qquad
			\theta_{\mathrm{dm}} = \frac{1}{2}  \left(k^2 \tau\right)
			\left[\frac{\Psi + \beta \delta \phi}{1 + \beta \tau \dot{\phi}/2}\right],
			\label{eq: thetaini}
			\\
			\sigma_{\nu} &= \frac{1}{15} \left( k \tau \right)^2 \Psi, \qquad
			\sigma_{\gamma} = \frac{\sigma_{\nu}}{1 + {9\dot{\tau}_c}{\tau}/20},
			\label{eq: sigmaini}
			\\
			F_{\nu3} &= \frac{2}{105} \left( k \tau \right)^3 \Psi, \qquad
			F_{\gamma3} = \frac{F_{\nu3}}{ \left( 1 + {\dot{\tau}_c}{\tau}/3 \right)}, 
			\label{eq: f3ini}
			\\
			\tau &= \frac{a}{H_0 \sqrt{\Omega_{\gamma}^{(0)} + \Omega_{\nu}^{(0)}}}, \qquad
			\Omega_{\mathrm{dm}} h^2 = \frac{\Omega_{\mathrm{dm}}^{(0)} h^2}{a^3},
			\label{eq: backgroundini}
		\end{align}
	\end{subequations}
	where $\beta = A,_{\phi}/A$, and $R_{\nu} = \rho_{\nu}/(\rho_{\nu} + \rho_{\gamma})$ is the neutrino-photon density ratio and $C$ is a normalization constant which sets the initial value of $\Psi$ (we choose $C = 0.5$ in accordance with its value in CAMB \cite{Lewis:1999bs}). One can check that with the absence of scalar field, these equations will reduce to the corresponding initial conditions in $\mathrm{\Lambda CDM}$ model.
	
	Eqs.~\eqref{eq: initial conditions} apply to a general modified gravity model with the scalar field non-minimally coupled to DM. What is missing is the choice for the initial value of the scalar field and its derivative, which need to be chosen based on the properties of a given model. Since we are interested in the symmetron model, in which the VEV of the scalar field is zero prior to symmetry breaking, we set $\phi$ to be a small nonzero number,  $\phi_{ini} \ll M_{pl}$, and $\dot{\phi}=\delta \phi=\delta \dot{\phi}=0$ at the initial time.

	\section{The quasi-static approximation} 
	\label{s:qsa}
	
	In what follows, we present the QSA form of the relevant perturbation equations.
	
	\subsection{QSA for the scalar field coupled to dark matter} 
	\label{ss:qsa dm}
	
	We revisit the equations of the previous section and apply the QSA. We take $k \gg \mathcal{H}$ and assume that 
	\begin{equation}
		\label{eq: qsa condition}
		k^2 \Phi \gg \left\{\ddot{\Phi}, \mathcal{H} \dot{\Phi}\right\}, \qquad
		k^2 \delta \phi \gg \left\{ \delta \ddot{\phi}, \mathcal{H} \delta \dot{\phi} \right\}.
	\end{equation}
	Under these approximations, the perturbation of the scalar field $\delta \phi$ is algebraically related to the DM matter density perturbation, and so are the gravitational potentials $\Phi$ and $\Psi$:
	\begin{subequations}
		\label{eq: qsa eq}
		\begin{align}
			\delta \phi &= - \frac{V'_{\mathrm{eff}} + \beta \rho_{\mathrm{dm}} \delta_{\mathrm{dm}}}{V''_{\mathrm{eff}} + k^2/a^2},
			\label{eq: deltaphi qsa}
			\\
			k^2 \Phi &= -4 \pi Ga^2 
			\left( 
			{\sum\limits_{i} } \rho_{i} \delta_{i}
			+ V,_{\phi} \delta{\phi} \right), 
			\label{eq:Phi qsa}
			\\
			k^2 \Psi &= k^2 \Phi -16 \pi G a^2 ( \rho_{\gamma} \sigma_{\gamma} + \rho_{\nu} \sigma_{\nu}),
			\label{eq: Psi qsa}
		\end{align}
	\end{subequations}  
	where $V'_{\mathrm{eff}}$  is given by Eq.~(\ref{eq: vprimeeff}).
Since QSA implies that the scalar field remains at the minimum of the effective potential, we have $V'_{\rm eff} = 0$ . Also, the contribution of the scalar field density perturbations, $V,_{\phi} \delta{\phi} $, to the Poisson equation is negligible compared to the matter density, hence we will omit it in Eq.~(\ref{eq:Phi qsa}). With this, we can rewrite Eqs.~\eqref{eq: qsa eq} as
	\begin{subequations}
		\label{eq: qsa mpc}
		\begin{align}
			\delta \phi &= -\beta \frac{3 \mathcal{H}_0^2 \Omega_{\mathrm{dm}} \delta_{\mathrm{dm}}}{m^2 + k^2/a^2},
			\label{eq: deltaphi qsa mpc}
			\\
			\Phi &= - \frac{3 \mathcal{H}_0^2 a^2}{2k^2} \left(
			\frac{\Omega_b^{(0)}}{a^3} \delta_b + \Omega_{\mathrm{dm}} \delta_{\mathrm{dm}} + \frac{\Omega_{\nu}^{(0)}}{a^4} \delta_{\nu} + \frac{\Omega_{\gamma}^{(0)}}{a^4} \delta_{\gamma} \right) ,
			%		- \frac{a^2 V,_{\phi} \delta \phi}{2 k^2},
			\label{eq:Phi qsa mpc}
			\\
			\Psi &= \Phi - \frac{6 \mathcal{H}_0^2 a^2}{k^2} \left( \frac{\Omega_{\nu}^{(0)}}{a^4}
			\sigma_{\nu} + \frac{\Omega_{\gamma}^{(0)}}{a^4} \sigma_{\gamma}
			\right),
			\label{eq:Psi qsa mpc}	
		\end{align}
	\end{subequations}
	where with times and distances are now in $\mathrm{Mpc}$ units and $m^2 = V''_{\mathrm{eff}}(\phi_{\rm min})$, with $\phi_{\rm min}$ being the minimum of $V_{\mathrm{eff}}$.
	
	\subsection{QSA for the scalar field coupled to all matter} 
	\label{ss:qsa all matter}
	
	Applying the QSA in the case of the scalar field coupled to all matter, the affected equations in the Einstein frame are
	\begin{subequations} \label{eq:qsa_all2}
	\renewcommand{\theequation}{\theparentequation.\arabic{equation}}
	\begin{align}
		&\mathcal{\tilde{H}}^2 = \frac{8\pi G \tilde{a}^2}{3} 
		\left(
		\tilde{\rho}_{\mathrm{dm}} + \tilde{\rho}_{b} + \tilde{\rho}_{\gamma} + \tilde{\rho}_{\nu} + \tilde{\rho}_{\Lambda}
		\right), \label{eq:fiedmann_lcdm_all2} \\
		&\tilde{\phi} = \tilde{\phi}_{\rm min}, \\
		&\tilde{\rho}_{\mathrm{dm}} = \frac{A}{A_0} \frac{\tilde{\rho}_{\mathrm{dm}}^{(0)}}{\tilde{a}^3}, \label{eq:rhodm_qsaall2} \\
		&\tilde{\rho}_b = \frac{A}{A_0} \frac{\tilde{\rho}_b^{(0)}}{\tilde{a}^3}, \label{eq:rhob_qsaall2} \\
		&k^2 \tilde{\Phi} = -4 \pi G\tilde{a}^2 \left( {\sum\limits_{i} } \tilde{\rho}_{i} \tilde{\delta}_{i} + \tilde{V},_{\tilde{\phi}} \delta \tilde{\phi} \right), \label{eq:cphi_qsaall2} \\
		&k^2 (\tilde{\Phi} - \tilde{\Psi}) = 16 \pi G \tilde{a}^2 ( \tilde{\rho}_{\gamma} \tilde{\sigma}_{\gamma} + \tilde{\rho}_{\nu} \tilde{\sigma}_{\nu}), \label{eq:shear_qsaall2} \\
		&\dot{\tilde{\delta}}_b + \tilde{\theta}_b = 0, \label{eq:deltadotb_qsaall2} \\
		&\dot{\tilde{\theta}}_b + \mathcal{\tilde{H}} \tilde{\theta}_b = k^2 \left(\tilde{\Psi} + \tilde{\beta} \delta \tilde{\phi} \right), \label{eq:thetadotb_qsaall2} \\
		&\dot{\tilde{\delta}}_{\mathrm{dm}} + \tilde{\theta}_{\mathrm{dm}} = 0, \label{eq:deltadotdm_qsaall2} \\
		&\dot{\tilde{\theta}}_{\mathrm{dm}} +  \mathcal{\tilde{H}} \tilde{\theta}_{\mathrm{dm}}  
		= k^2 \left( \tilde{\Psi} + \tilde{\beta} \delta \tilde{\phi} \right), \label{eq:thetadotdm_qsaall2} \\
		&\delta \tilde{\phi} = - \frac{\tilde{\beta} \left( \tilde{\rho}_{\mathrm{dm}} \tilde{\delta}_{\mathrm{dm}} + \tilde{\rho}_{b} \tilde{\delta}_{b} \right)}{\tilde{m}^2 + k^2/\tilde{a}^2}, \label{eq:deltaphi_qsaall2} \\
		&\tilde{m}^2 = \tilde{V}''_{\mathrm{eff}}(\tilde{\phi}_{\rm min}). \label{eq: vdprimeeff_ein}
	\end{align}
	\end{subequations}
	Transforming these equations into the baryon frame gives Eqs.~\eqref{eq:qsa_all}, where we have omitted the $\tilde{V},_{\tilde{\phi}} \delta \tilde{\phi}$ term in \eqref{eq:cphi_qsaall2} due to its negligible effect.
	
	\section{Parameters of the Symmetron model} \label{app:parameters symmetron}
	In this section, we provide the relations between the parameters appearing in the Lagrangian of the symmetron model and the phenomenological parameters $(a_{SSB}, \lambda_C, \beta_0)$ used in the text.
	We start by defining two parameters which set the coupling strength at zero matter density and at present time, respectively:
	\begin{align}
		\beta_* &= \frac{\phi_*}{M^2}, \label{eq:betastar}
		\\
		\beta_0 &= \frac{\phi_0}{M^2} = \beta_* \sqrt{1 - a_{SSB}^3}, \label{eq:beta0}
	\end{align}
	where $\phi_* \equiv \mu / \sqrt{\lambda}$ is the expectation value of scalar field at zero matter density, and we have used Eq.~\eqref{eq:phi qsa} to derive the second expression. Eq.~\eqref{eq:m_qsa} for the effective mass is derived by substituting the value of $\phi$ at the minimum of the effective potential, given by Eq.~\eqref{eq:phi qsa}, into $V''_{\rm eff}$ for the symmetron model, given by
\begin{equation}
	V''_{\rm eff}(\phi_{\rm min}) = 3\lambda \phi_{\rm min}^2 - \mu^2 \left( 1 - \frac{a_{SSB}^3}{a^3} \right),
	\label{eq:vdpeff_min} 
\end{equation}
where we have used Eqs.~\eqref{eq:veff symmetron}, \eqref{eq:a_ssb} to derive \eqref{eq:vdpeff_min}. This leads to
	\begin{equation}
		m = m_* \sqrt{1 - \left( {a_{SSB}}/{a} \right)^3 },
		\label{eq:meff qsa}
	\end{equation}
	where $m_* = \sqrt{2} \mu = 1/ \lambda_*$ is the inverse of the Compton wavelength of the scalar field at zero matter density. Similarly to $\beta_0$, we can define $\lambda_C$ as the present value of Compton wavelength and relate the parameter $\lambda_*$ to $\lambda_C$ as
	\begin{equation}
		\lambda_C = \frac{\lambda_*}{\sqrt{1 - a_{SSB}^3}}. \label{eq:lambda0}
	\end{equation}
To find the relation for the remaining parameter, $M$, we start from the condition at symmetry breaking given by Eq.~\eqref{eq:a_ssb} and define a new parameter $M_s^2$ as
	\begin{equation}
		M_s^2 = \frac{\rho_{\mathrm{dm}}^{(0)}}{\mu^2 a_{SSB}^3}. \label{eq:m_s}
	\end{equation}
	Next, we use Eqs.~\eqref{eq:a_ssb}, \eqref{eq:beta0} to derive a quartic equation for $M$ given by
	\begin{equation}
		\frac{1}{2} \beta_0^2 M^4 + M^2 - M_s^2 = 0. \label{eq:quartic M}
	\end{equation}
	This equation has one real positive solution which can be expressed as
	\begin{equation}
		M = \frac{1}{\beta_0} \sqrt{\left[\sqrt{1 + 2 M_s^2 \beta_0^2} -1 \right]}.
	\end{equation}
The expressions for the phenomenological parameters $(a_{SSB}, \lambda_C, \beta_0)$ can now be written as
	\begin{align}
		a_{SSB}^3 &= \left( \frac{1}{2} + \frac{\lambda M^2}{\mu^2} \right) - \sqrt{\left( \frac{1}{2} + \frac{\lambda M^2}{\mu^2} \right)^2 - \frac{2\lambda \rho_{\mathrm{dm}}^{(0)}}{\mu^4}} , \label{eq:a_ssb_orig}
		\\
		\lambda_C^2 &= \frac{1}{\left( \mu^2 - \lambda M^2 \right) + \sqrt{\left( \mu^2 + \lambda M^2 \right)^2 -8\lambda \rho_{\mathrm{dm}}^{(0)}}} , \label{eq:lambdac_orig}
		\\
		\beta_0^2 &= \left( \frac{\mu^2}{2 \lambda M^4} - \frac{1}{M^2} \right) + \sqrt{\left( \frac{\mu^2}{2 \lambda M^4} + \frac{1}{M^2} \right)^2 - \frac{2 \rho_{\mathrm{dm}}^{(0)}}{\lambda^2 M^8}} . \label{eq:beta0_orig}
	\end{align}
	One can simply check that for the case of zero DM density corresponding to the vacuum solution of the symmetron model, we have $(\lambda_C, \beta_0) \rightarrow (\lambda_*, \beta_*)$ consistent with Eqs.~\eqref{eq:betastar}, \eqref{eq:lambda0}.

\end{document}